\documentclass[fleqn,usenatbib]{mnras}

\usepackage{newtxtext,newtxmath}

\usepackage[T1]{fontenc}

\DeclareRobustCommand{\VAN}[3]{#2}
\let\VANthebibliography\thebibliography
\def\thebibliography{\DeclareRobustCommand{\VAN}[3]{##3}\VANthebibliography}

\usepackage{mystyle}

\newcommand{\ngc}[1]{\ifthenelse{\equal{#1}{}}{NGC~3256}{NGC~#1}}
\makeglossaries
\newacronym{agn}{AGN}{active galactic nucleus}
\newacronym{aca}{ACA}{Atacama Compact Array}
\newacronym{alma}{ALMA}{Atacama Large Millimeter/Submillimeter Array}
\newacronym[firstplural=cluster-formation efficiencies (CFEs)]{cfe}{CFE}{cluster-formation efficiency}
\newacronym{co}{CO}{carbon monoxide}
\newacronym{casa}{\textsc{casa}}{Common Astronomy Software Applications}
\newacronym{ci}{CI}{confidence interval}
\newacronym{cmb}{CMB}{cosmic microwave background}
\newacronym{carta}{\textsc{carta}}{Cube Analysis and Rendering Tool for Astronomy}
\newacronym{dec}{Dec.}{declination}
\newacronym{fov}{FoV}{field of view}
\newacronym{fwhm}{FWHM}{full width at half maximum}
\newacronym{gmc}{GMC}{giant molecular cloud}
\newacronym{hwhm}{HWHM}{half width at half maximum}
\newacronym{ir}{IR}{infrared}
\newacronym{idl}{\textsc{IDL}}{Interactive Data Language}
\newacronym{ism}{ISM}{interstellar medium}
\newacronym{kde}{KDE}{kernel density estimator}
\newacronym{ks}{KS}{Kolmogorov-Smirnov}
\newacronym{lirg}{LIRG}{luminous infrared galaxy}
\newacronym{mad}{MAD}{median absolute deviation}
\newacronym{ned}{NED}{NASA/IPAC Extragalactic Database}
\newacronym{phangs}{PHANGS-ALMA}{Physics at High Angular resolution in Nearby GalaxieS with ALMA}
\newacronym{pa}{PA}{position angle}
\newacronym{ppv}{PPV}{position-position-velocity}
\newacronym{ra}{R.A.}{right ascension}
\newacronym{rms}{RMS}{root-mean-square}
\newacronym{s/n}{S/N}{signal-to-noise}
\newacronym{spw}{SPW}{spectral window}
\newacronym{sfr}{SFR}{star-formation rate}
\newacronym{tp}{TP}{total power}
\newacronym{u/lirg}{U/LIRG}{ultra/luminous infrared galaxy}
\newacronym{wmap}{\emph{WMAP}}{Wilkinson Microwave Anisotropy Probe}

\title[Extreme GMCs in NGC~\num{3256}]{Extreme giant molecular clouds in the luminous infrared galaxy NGC~\num{3256}}

\author[N. Brunetti and C. D. Wilson]{
Nathan Brunetti\thanks{E-mail: brunettn@mcmaster.ca} and
Christine D. Wilson
\\
Department of Physics and Astronomy, McMaster University, Hamilton, ON L8S 4M1, Canada
}

\date{Accepted XXX. Received YYY; in original form ZZZ}

\pubyear{2022}

\begin{document}
\label{firstpage}
\pagerange{\pageref{firstpage}--\pageref{lastpage}}
\maketitle

\begin{abstract}
We present a cloud decomposition of $^{12}$CO (\num{2}--\num{1}) observations of the merger and nearest luminous infrared galaxy, \ngc{}.
\num{185} spatially and spectrally resolved clouds are identified across the central \SI{\approx 130}{\square\kilo\parsec} at \SI{90}{\parsec} resolution and completeness is estimated.
We compare our cloud catalogue from \ngc{} to ten galaxies observed in the \acrshort{phangs} survey.
Distributions in \ngc{} of cloud velocity dispersions (median \SI{23}{\kilo\metre\per\second}), luminosities (\SI{1.5e7}{\kelvin\kilo\metre\per\second\square\parsec}), CO-estimated masses (\SI{2.1e7}{\solarmass}), mass surface densities (\SI{470}{\solarmass\per\square\parsec}), virial masses (\SI{5.4e7}{\solarmass}), virial parameters (\num{4.3}), size-linewidth coefficients (\SI{6.3}{\square\kilo\metre\per\square\second\per\parsec}), and internal turbulent pressures (\SI{1.e7}{\kelvin\per\cubic\centi\metre}$\, k_{\mathrm{B}}^{-1}$) are significantly higher than in the \acrshort{phangs} galaxies.
Cloud radii (median \SI{88}{\parsec}) are slightly larger in \ngc{} and free-fall times (\SI{4.1}{\mega\year}) are shorter.
The distribution of cloud eccentricities in \ngc{} (median of \num{0.8}) is indistinguishable from many \acrshort{phangs} galaxies, possibly because the dynamical state of clouds in \ngc{} is similar to that of nearby spiral galaxies.
However, the narrower distribution of virial parameters in \ngc{} may reflect a narrower range of dynamical states than in \acrshort{phangs} galaxies.
No clear picture of cloud alignment is detected, despite the large eccentricities.
Correlations between cloud properties point to high external pressures in \ngc{} keeping clouds bound and collapsing given such high velocity dispersions and star-formation rates.
A fit to the cloud mass function gives a high-mass power-law slope of $-2.75^{+0.07}_{-0.01}$, near the average from \acrshort{phangs} galaxies.
We also compare our results to a pixel-based analysis of these observations and find molecular-gas properties agree qualitatively, though peak brightness temperatures are somewhat higher and virial parameters and free-fall times are somewhat lower in this cloud-based analysis.
%The abstract should briefly describe the aims, methods, and main results of the paper.
%It should be a single paragraph not more than 250 words (200 words for Letters).
%No references should appear in the abstract.
\end{abstract}


%%%%%%%%%%%%%%%%%%%%%%%%%%%%%%%%%%%%%%%%%%%%%%%

%%%%%%%%%%%%%%%%% BODY OF PAPER %%%%%%%%%%%%%%%%%%

\section{Introduction}
Understanding the conditions which give rise to varying levels of star formation is an important part of fully describing galaxy evolution over cosmic time.
Stellar feedback has dramatic morphological, kinematic, thermodynamic, and chemical impacts on the host galaxy, especially the \gls{ism}.
The injection of energy by stars into their surroundings contributes to the turbulent energy that supports the gas disc against gravity, setting the scale height of the atomic and molecular \gls{ism} \citep[e.g.][]{McK1977, Nar2002, Ost2011, Elm2011, Ben2016, Kru2018}.
Stellar radiation heats up pockets of the \gls{ism} around centres of active star formation through the far-ultraviolet photoelectric effect on polycyclic aromatic hydrocarbons and dust grains \citep{Tie1985, McK1989, Hol1999}.
Mechanically driven shocks from stellar winds and supernovae also contribute to heating the surrounding \gls{ism} \citep{McK1977, Kru2006, Hop2012}.
It has been widely shown that the typical rate of star formation within galaxies has dramatically changed over the age of the universe \citep[see e.g. the review by][]{Mad2014}.
As the \gls{sfr} in a galaxy changes throughout its life, the dominance of star formation in driving morphological, kinemtatic, thermodynamic, and chemical changes will also evolve.

Since the majority of stars form in molecular gas, we focus on exploring the link between the conditions of the molecular gas and the resulting \gls{sfr}.
In particular, we aim to compare the molecular gas properties in nearby spiral galaxies with those within nearby merging galaxies.
Mergers cause galaxies to enter a starburst phase \citep{Lar1978, Ell2008, Ell2013, Han2020} in which they exhibit elevated \glspl{sfr} relative to spiral galaxies as well as relative to their total amount of molecular gas \citep{Dad2010, Yam2017, Her2019, Wil2019, Ken2021}.
Therefore, comparing mergers to spiral galaxies in detail will allow us to explore the conditions of the molecular \gls{ism} that lead to more vigorous modes of star formation.

Merging systems may also recreate some of the conditions for star formation that existed in high-redshift galaxies.
Local merging galaxies have \glspl{sfr} that are similar to those in high-redshift galaxies \citep{Zar2018, Lar2020, Elm2021, Elm2009}. %Elm2009 compared the \glspl{sfr} of clumps in local flocculent spirals and dwarf irregulars to high-redshift galaxies, and the rest were comparisons of \glspl{sfr} in local mergers to non-interacting local galaxies and/or high-redshift galaxies.
Additionally, the galaxy merger fraction increases with redshift \citep[e.g. see the compilation and comparison to literature merger fraction estimates spanning $z \lesssim \num{6}$ by][]{Rom2021}.
These similarities mean that since we can observe much smaller physical scales in nearby mergers than in high-redshift galaxies, we should be able to link the smaller-scale properties from nearby mergers to the larger scales at high redshift to build a more complete picture of star formation and \gls{ism} properties over cosmic time \citep[for a review of the progress made in measuring the properties of the \gls{ism} in high-redshift systems see][]{Tac2020}.

At a distance of \SI{44}{\mega\parsec} (Table~\ref{tab:fits2props_arguments}), \ngc{} is the nearest \gls{lirg}, forming stars at a rate of about \SI{50}{\solarmass\per\year} \citep{Sak2014}.
It is a late-stage merger where the progenitor galaxies' nuclei are still separated but share a common envelope of gas, dust, and stars \citep{Sti2013}.
\ngc{} also provides, within a single system, a fairly diverse set of environments in which to study molecular gas.
The orientation of the northern nucleus appears to be nearly face-on, while the southern nucleus is nearly edge-on \citep[for diagrams see figures \num{18} and \num{1} from][respectively]{Sak2014, Har2018}.
This edge-on orientation offers a clear view of the spectacular bipolar jet that extends \SI{700}{\parsec} north and south of the southern nucleus while only being about \SI{140}{\parsec} wide \citep{Sak2014}.
An analysis by \citet{Sak2014} of the energy required to drive the jet indicates it must be at least partially driven by an \gls{agn}, consistent with \gls{ir} and X-ray observations \citep{Ohy2015}.
There is also evidence for a red and blueshifted starburst-powered outflow being launched from the northern nucleus at an angle almost parallel with the line of sight \citep{Sak2014}.

In this paper we compare clouds identified in \gls{co} $J$=\num{2}--\num{1} observations of \ngc{} with clouds found in ten nearby spiral galaxies observed by the \gls{phangs} survey \citep{Ros2021}.
Details of the observations, imaging, and data preparation are summarized in Section~\ref{sec:data}.
The cloud-finding procedure is described in Section~\ref{sec:analysis}, as well as the characterization of cloud-finding completeness.
Section~\ref{sec:results} covers the main results from comparing our cloud catalogue from \ngc{} to those from \gls{phangs}, and in Section~\ref{sec:discussion} we discuss physical implications of the similarities and differences as well as compare to a pixel-based analysis presented in \citet{Bru2021}.
Finally, we summarize our results and conclusions in Section~\ref{sec:conclusions}.

\section{Data} \label{sec:data}
\subsection{Observations}
Observations of the \gls{co} (\num{2}--\num{1}) emission line were carried out towards \ngc{} with the \gls{alma} main array, the \gls{aca}, and \gls{tp} array.
See \citet{Bru2021} for a complete description of the observations, calibration, and imaging of the interferometric data.
The results were cubes covering the central \SI{6.4}{\kilo\parsec} radius of \ngc{} (or about \SIrange{130}{140}{\square\kilo\parsec}) from \SIrange{2004}{3599}{\kilo\metre\per\second} with \SI{5.131}{\kilo\metre\per\second} wide channels (using the radio convention) with a synthesized beam \gls{fwhm} of \num{0.25} arcsec (\SI{\sim 53}{\parsec}).

To capture the high velocity outflow and jet present in \ngc{}, the spectral setup of the telescope was arranged such that the edges of two \glspl{spw} covered the \gls{co} line to provide a relatively wide bandwidth with high spectral resolution.
A fraction of the \glspl{spw} overlap in frequency, occurring over the central portion of the spectral line.
When reducing the \gls{tp} observations with the \gls{alma} pipeline, this spectral setup resulted in the two \glspl{spw} producing discontinuities in the combined spectra.
For this reason the single-dish observations were not combined with the inteferometric data in the first paper.
 
We have since re-attempted reducing the \gls{tp} observations with the \gls{phangs} \gls{tp} pipeline\footnote{
    The scripts can be found at \url{https://github.com/PhangsTeam/TP_ALMA_data_reduction}.
    Accessed 2020 July 14.
    A modified version made to handle multiple \glspl{spw} for a single spectral line will be shared on reasonable request to the corresponding author.
}.
While this produced a cube of the combined \glspl{spw} without obvious discontinuities along the spectra, imaging the two \glspl{spw} separately still revealed a maximum difference between beam-averaged intensities of \num{41} per cent.
We estimated the fraction of recovered flux in the original interferometric observations is \num{90} per cent compared to a cube produced by feathering the interferometric and \gls{tp} cubes in the \gls{casa}.
Given the large difference between the two \glspl{spw} that make up the \gls{tp} cube, and the high recovered-flux fraction in the interferometric cube, we again chose to carry out this analysis on the interferometric-only observations.
As a result, the total mass will be underestimated by at most ten per cent, and this bias will be strongest for the spatially-largest clouds.

\subsection{Matching resolution and producing uniform-noise cubes} \label{sec:matching_beams_and_noise}
There have been many studies on the biasing effects of differing resolutions and noise levels on cloud decompositions \citep{Rei2010, Hug2013, Ros2021}.
These studies clearly show that mitigating those differences is essential to making robust comparisons between data sets.
Since our primary comparison in this work is to the homogenized cloud catalogue from \citet{Ros2021} we focus on replicating their procedures as closely as possible.
The first step was to correct the cleaned \gls{co} cube for the primary beam response.
We then convolved this cube to have a synthesized beam \gls{fwhm} of \SI{90}{\parsec}.

To simplify the effects that the noise level in the data has on the recovered mass fraction of each cloud and the cloud-finding completeness, we produced uniform-noise \gls{co} cubes following the procedure described by \citet{Ros2021}.
Note that we made noise estimates for each pixel in spectral and spatial dimensions following the procedure and parameters used by \citet{Sun2018}\footnote{
    The \textsc{Python} script for producing noise cubes was obtained from \url{https://github.com/astrojysun/Sun_Astro_Tools/blob/master/sun_astro_tools/spectralcube.py} and we used the version at commit f444343.
}
and \citet{Bru2021}, not the PHANGS-ALMA pipeline used by \citet{Ros2021}.

While our observations of \ngc{} were roughly the same sensitivity as those of the \gls{phangs} sample in \si{\jansky\per\beam}, the greater distance meant we had about ten times higher noise in the Kelvin scale.
Since we could not match the noise level of \citet{Ros2021} (\SI{0.075}{\kelvin} across all galaxies after noise homogenization), we tested several noise-level targets and chose to proceed with a value of \SI{0.9}{\kelvin} as it was the lowest noise value that would not exclude clouds due to reducing the usable \gls{fov}.
We also note that  the observations of \ngc{} have channels that are about a factor of two wider than the \gls{phangs} observations.
We will need to be cautious when comparing \ngc{} and \gls{phangs} data sets due to the different noise levels and channel widths.

\section{Analysis} \label{sec:analysis}
\subsection{Cloud finding} \label{sec:cloud_finding}
To identify discrete molecular-gas structures in \ngc{} we used the \textsc{pycprops}\footnote{\url{https://github.com/PhangsTeam/pycprops}} \textsc{Python} package \citep{Ros2021}.
This package is a translation of the original \gls{idl} \textsc{cprops} package \citep{Ros2006}, that takes advantage of the \textsc{astrodendro}\footnote{\url{http://www.dendrograms.org/}} package for segmenting emission to increase speed as well as other changes to improve robustness for comparative analyses.
\citet{Ros2021} provide a detailed description of the algorithm, so we focus on our parameter choices and how they compare to \citet{Ros2021}.

For the most direct comparison, we use the same criteria for finding maxima as \citet{Ros2021}, which are \num{1}) a minimum contrast between maxima of two times the noise in the cube, \num{2}) a minimum number of pixels that corresponds to one quarter of the beam solid angle, \num{3}) no minimum separation between maxima, and \num{4}) no minimum change in properties from merging maxima.
We also set the `compactness' parameter to the same value as \citet{Ros2021}, such that pixels were assigned to produce the most compact structures by the watershed algorithm.
Specific arguments to the \textit{fits2props} function are shown in Table~\ref{tab:fits2props_arguments}.

\begin{table}
    \centering
    \caption{
        Arguments to \textit{fits2props}.
    }
    \begin{threeparttable}
        \label{tab:fits2props_arguments}
        \begin{tabular}{
            @{}
            l
            r
            @{}
        }
            \toprule
            Argument name & Value \\
            \midrule
            `distance'    & \SI{44}{\mega\parsec}{\tnote{a}} \\
            `alphaCO'     & \SI{1.38}{\solarmass\per\square\parsec}(\si{\kelvin\kilo\metre\per\second})$^{-1}$ \\
            `channelcorr' & \num{0.185} \\
            `minpix'      & \num{29} \\
            `sigdiscont'  & \num{0} \\
            `compactness' & \num{1000} \\
            `specfriends' & \num{0} \\
            `friends'     & \num{0} \\
            `rmstorad'    & $\sqrt{2 \ln{2}}$ \\
            `bootstrap'   & \num{100} \\
            \bottomrule
        \end{tabular}
        \begin{tablenotes}
            \item [a] \acrshort{cmb}-corrected redshift retrieved from \acrshort{ned}, using \acrshort{wmap} five-year cosmology with $H_{0} = \SI{70.5}{\kilo\metre\per\second\per\mega\parsec}$, $\Omega = \num{1}$, and $\Omega_{\mathrm{m}} = \num{0.27}$.
            \item [] \textit{Notes.}  Used version \href{https://github.com/PhangsTeam/pycprops/commit/1462ff416389de9d2cea80c5f08979e470d55f99}{1462ff4} of \textsc{pycprops}.
            Arguments not listed here were left as their default values.
        \end{tablenotes}
    \end{threeparttable}
\end{table}

Despite attempting to flatten the noise throughout the \gls{co} cube to \SI{0.9}{\kelvin}, as described in Section~\ref{sec:matching_beams_and_noise}, we still calculated a noise cube from the flattened cube to account for any noise variations still present in the data.
The median of all pixels in the noise cube was calculated to be \SI{0.897}{\kelvin} and the minimum contrast between maxima was set to the default of two times this median noise.
We also chose to use a signal mask when running \textsc{pycprops} to limit the pixels assigned to maxima to only those with likely significant emission.
This signal mask was produced from the flattened-noise cube with the same script used to make the noise cubes.
We note that the procedure to generate the noise and signal-mask cubes was not identical to the \gls{phangs} pipeline, used by \citet{Ros2021}, but they were designed by \citet{Sun2018} to be used for the same purposes \citetext{E. Rosolowsky, private communication}.

\textsc{pycprops} requires an estimate of the channel-to-channel correlation to remove the finite channel response from the velocity dispersions of the clouds.
We estimate this correlation from emission-free channels with the Pearson correlation coefficient calculated from all pixel values between the $i$ and $i + 1$ channels (the `channelcorr' argument).
Internally, \textsc{pycprops} estimates major and minor axis lengths for each cloud and combines these into a single radius for each structure, $R$.
This radius is calculated as $R = \eta \sqrt{\sigma_{\mathrm{maj,d}}\sigma_{\mathrm{min,d}}}$ \citep[equation 9 from][]{Ros2021} where $\sigma_{\mathrm{maj,d}}$ and $\sigma_{\mathrm{min,d}}$ are the major and minor sizes, assumed to be the spatial standard deviations of Gaussian clouds, and $\eta$ is a factor that depends on the mass distribution within the cloud.
$\eta$ corresponds to the `rmstorad' argument to \textit{fits2prop} in Table~\ref{tab:fits2props_arguments} and we have adopted the same value of $\sqrt{2 \ln{2}} \approx 1.18$ as used by \citet{Ros2021}, corresponding to a Gaussian density profile and assuming $R$ is the \gls{hwhm} of the cloud.

Cloud finding was carried out at \SI{90}{\parsec} resolution to simplify comparisons to results from \citet{Ros2021}, and this resolution can also be roughly compared to the \SI{80}{\parsec} resolution pixel-based analysis from \citet{Bru2021}.
However, instead of using cubes regridded to have pixels that were half the beam \gls{fwhm}, as done by \citet{Bru2021}, we kept the original pixel grid from the original cleaned cube.
This was because we found an unexpected trend of increasing cloud velocity dispersion distributions with decreasing pixel size.
See Appendix~\ref{sec:dispersion_pixel_size} for details of the tests which showed this behaviour.

\subsection{Estimating source-finding completeness} \label{sec:completeness}
Interpretation of the cloud-property distributions measured in \ngc{} depends on understanding the impact that the noise level, resolution, and choice of cloud-finding algorithm have on the types of clouds we are able to find.
To empirically infer the completeness limits on our cloud finding we performed Monte Carlo tests by injecting \num{1200} synthetic sources with unique properties into our data, one at a time, and attempted to find them with the same steps described in Section~\ref{sec:cloud_finding}.
We recorded if the injected sources were found as well as the cloud properties estimated by \textsc{pycprops}.

We chose to inject sources into a subset of channels at the high-velocity end of the original emission cube that did not contain significant real emission.
This process simplified the interpretation of the source-finding results since any clouds found would be attributed to the injection of the synthetic source.
However, a limitation is that the effect of source blending is not incorporated in our completeness estimate.
Blending of clouds likely plays a non-trivial role as most of the emission in \ngc{} is interconnected through the cube with few instances of isolated ``islands'' of emission.

It is worth noting that the procedure for finding synthetic clouds was not exactly identical to cloud finding in the real emission.
The difference was that instead of recalculating the noise cube after injecting each source, we used the noise cube from the original source finding, trimmed to the same channels as the emission cube.
We found that recalculating the noise cube from just that subset of channels resulted in subtle noise structures across the \gls{fov} not appearing in the resulting noise cube.
These structures appeared to be originating from channels with significant emission, which explained why they were missing in the recalculated synthetic-source noise cube.
We chose to preserve the effects that the real emission was having on the noise cube over any that would be introduced when the synthetic sources were injected.

To choose the synthetic source properties, we followed \citet{Ros2021} by uniformly sampling (in log-space) cloud masses, surface densities, and virial parameters.
The distributions of these properties were centred on \SI{2e7}{\solarmass}, \SI{600}{\solarmass\per\square\parsec}, and \num{7} with widths of \SIlist{2.75; 2; 2.25}{\dex}, respectively.
The ranges of synthetic properties were chosen such that the upper limit is about two times the largest value from our cloud catalogue and the lower limit is about half the smallest.
Masses, surface densities, and virial parameters were used to calculate three-dimensional Gaussian parameters corresponding to the spatial and velocity standard deviations.
These Gaussian parameters were then added, in quadrature, to the beam standard deviation or equivalent Gaussian channel width \citep[$\sigma_{v,\mathrm{chan}}$ in equation 6 from][]{Ros2021} to replicate the effects of resolution on the synthetic clouds' true properties.
Peak brightness temperatures were then calculated from the mass and ``convolved'' spatial and spectral sizes still following the three-dimensional Gaussian cloud model of \citet{Ros2021}.
\Gls{ppv} Gaussians were calculated from these parameters and were added to the emission cube.

For each of the \num{1200} choices of properties we also chose to separately inject and search for five identical sources that were centred at different \gls{ra}, \gls{dec}, and velocity positions.
By injecting identical sources at different locations we reduce the stochastic nature of sources being more easily found (missed) if placed on a noise peak (trough).
Five positions were uniformly drawn from ranges of \gls{ra}, \gls{dec}, and velocity, and each choice of cloud properties was injected at those five locations.
The \gls{ra} and \gls{dec} ranges were limited to a region whose centre coincides with the centre of the \gls{fov}, with a width of \SI{7.25}{\kilo\parsec} in \gls{ra} and height of \SI{4.5}{\kilo\parsec} in \gls{dec}.
Placing the synthetic sources within this region ensured the centres of the largest synthetic clouds were at least two spatial standard deviations from the mapped edges.
Similarly, the number of channels that made up the subset of the cube in which synthetic sources were injected was chosen to cover a velocity range corresponding to $\sim 1 \sigma_{v}$ for the largest synthetic cloud velocity width chosen, and the velocity centres were chosen to be \num{\sim 0.4} velocity standard deviations from the edge channels.

\begin{figure}
    \centering
    \includegraphics{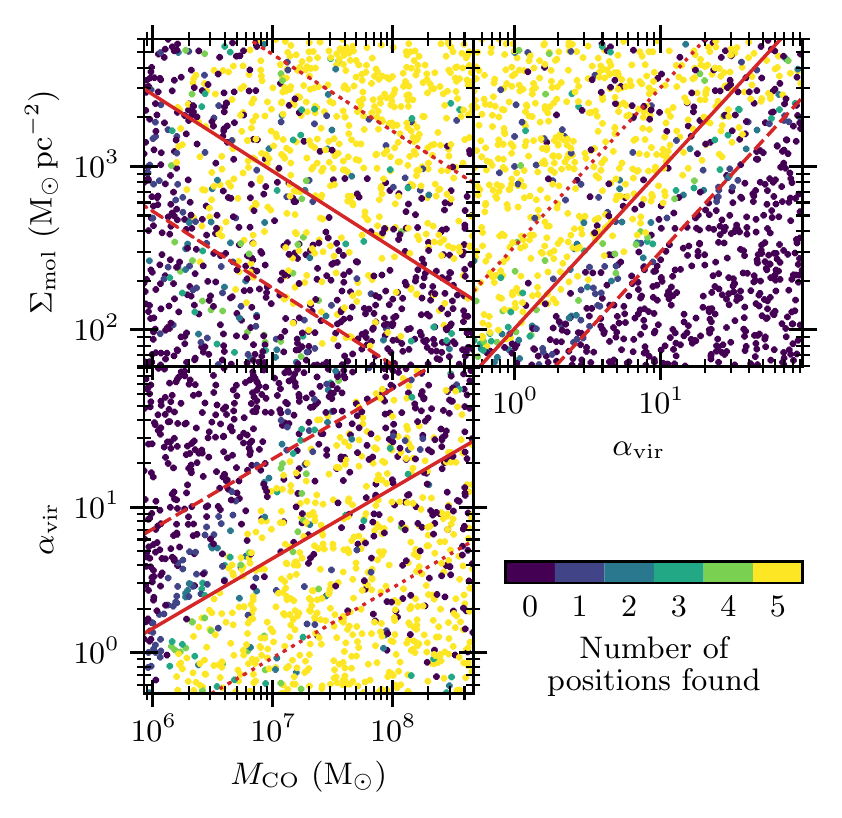}
    \caption{
        Results of Monte Carlo completeness tests from injecting synthetic sources with \num{1200} unique cloud properties chosen from log-uniform distributions of mass, surface density, and virial parameter.
        Each point represents one choice of cloud properties.
        Each choice of properties was injected five separate times into the cube at different \gls{ra}, \gls{dec}, velocity positions and \textsc{pycprops} was run to attempt to find each source.
        The colours of the points indicate how many of the five different positions had at least one cloud found by \textsc{pycprops}.
        Red lines indicate the predicted \num{80} (dotted), \num{50} (solid), and \num{20} (dashed) per cent completeness contours after averaging the three-dimensional fit along the axis not shown in each of the panels.
    }
    \label{fig:completeness_results_original}
\end{figure}

\begin{figure}
    \centering
    \includegraphics{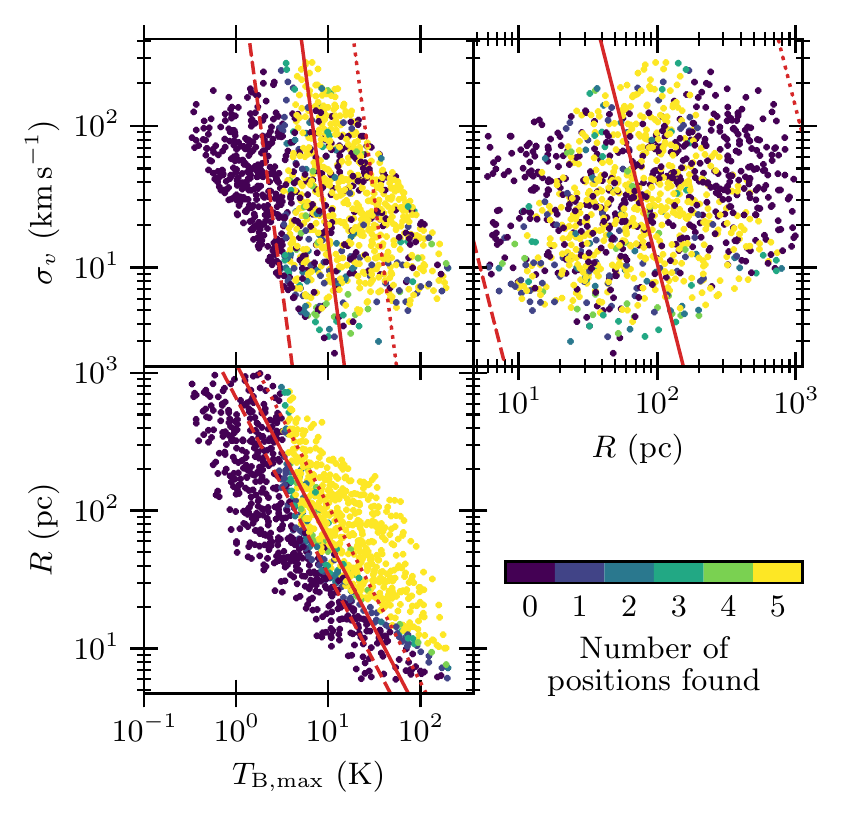}
    \caption{
        Same as Figure~\ref{fig:completeness_results_original} but with the cloud properties converted to the observable properties of peak brightness temperature, velocity dispersion, and radius.
        To calculate the Gaussian \gls{ppv} synthetic sources injected for the completeness tests, these values of $R$ and $\sigma_{v}$ are first combined with the beam and channel width, respectively, to emulate observational effects.
        Then the $T_{\mathrm{B,max}}$ values are recalculated from these ``observed'' $R$ and $\sigma_{v}$ values, and finally the Gaussian source is calculated from these properties.
        The velocity dispersion vs. radius panel appears to have its fitted completeness trend rising in the opposite direction to the scatter-point colours; see discussion in Section~\ref{sec:completeness}.
    }
    \label{fig:completeness_results_Gaussian}
\end{figure}

Figures~\ref{fig:completeness_results_original} and \ref{fig:completeness_results_Gaussian} summarize the results of our completeness estimation.
Each point in Figure~\ref{fig:completeness_results_original} represents a combination of cloud properties chosen in the mass-surface density-virial parameter space.
In Figure~\ref{fig:completeness_results_Gaussian} the results are shown converted to the observable peak brightness temperature-velocity dispersion-2D radius cloud property space (before applying the effects of resolution).
The colours of the points indicate how many of the five unique positions were found by \textsc{pycprops} such that regions of high and low completeness are the yellow and purple regions, respectively.
The coupled nature of the properties shown in Figure~\ref{fig:completeness_results_original} is apparent as no single property, or even pair of properties, straightforwardly determines the likelihood a cloud will be found.
The observational constraints are clearer in Figure~\ref{fig:completeness_results_Gaussian}, primarily in the peak brightness temperature-radius plane, where the likelihood of a cloud being found depends strongly on the part of parameter space in which it appears.
The boundary between detectable and not-detectable clouds is similarly sharp in the mass-surface density-virial parameter space but is not as clear when the results are projected into the panels of Figure~\ref{fig:completeness_results_original}.

Following \citet{Ros2021}, we perform a binomial logistic regression through maximum likelihood estimation to fit the completeness results with the function
\begin{alignat}{3} \label{eq:logistic}
    P(M, \Sigma, \alpha_{\mathrm{vir}}) = &\left\{ \right.& \\
                                          &               & 1 + \exp \left[ \right.\nonumber \\
                                          &               &                &- c_{0}\nonumber \\
                                          &               &                &- c_{1} \log_{10}{\left(\frac{M}{\SI{1e6}{\solarmass}}\right)}\nonumber \\
                                          &               &                &- c_{2} \log_{10}{\left(\frac{\Sigma}{\SI{150}{\solarmass\per\square\parsec}}\right)}\nonumber \\
                                          &               &                &- c{_3} \log_{10}{\left(\frac{\alpha_{\mathrm{vir}}}{2}\right)}\nonumber \\
                                          &               &\left. \right]\nonumber \\
                                          &\left. \right\}^{-1}\nonumber
\end{alignat}
with the best-fitting parameters and 95 per cent \glspl{ci} summarized in Table~\ref{tab:completeness_parameters}.
To be clear, the fit is not done to the fraction of positions detected for each choice of cloud properties.
Instead, for each combination of mass, surface density, and virial parameter (the independent variables) the dependent variable is either zero to mark no clouds found or one to mark at least one cloud found.
With a functional form for the completeness we calculated a grid of completeness predictions over our range of synthetic source properties and plot the \numlist{80; 50; 20} per cent completeness contours in Figures~\ref{fig:completeness_results_original} and \ref{fig:completeness_results_Gaussian}, averaged along the third axis not shown in each panel.

The velocity dispersion vs. radius panel in Figure~\ref{fig:completeness_results_Gaussian} appears to have its fitted completeness trend rising in the opposite direction to the scatter-point colours.
However, this is due to different ranges of velocity dispersion and radius being probed by our completeness tests at different values of peak brightness temperature.
If the scatter points are limited to a narrow range of temperatures then both the points and the fit trend show generally lower completeness for combined small radii and velocity dispersions and high completeness for large radii with high velocity dispersions.
Ultimately, the combination of velocity dispersion and radius alone is not a strong predictor of completeness.

\begin{table}
    \centering
    \caption{
        Results of logistic regression fit to completeness results. Estimated parameters correspond to those in Equation~\ref{eq:logistic}.
    }
    \begin{threeparttable}
        \label{tab:completeness_parameters}
        \begin{tabular}{
            @{}
            l
            S[table-format=+1.1]
            S[table-format=+1.1]
            S[table-format=+1.1]
            @{}
        }
            \toprule
            Parameter & {Best-fitting value} & {Lower 95\% \gls{ci}} & {Upper 95\% \gls{ci}}  \\
            \midrule
            $c_{0}$   & -2.8                 & -3.1                  & -2.6 \\
            $c_{1}$   & 1.8                  & 1.7                   & 2.0 \\
            $c_{2}$   & 3.9                  & 3.7                   & 4.1 \\
            $c_{3}$   & -3.8                 & -4.0                  & -3.6 \\
            \bottomrule
        \end{tabular}
    \end{threeparttable}
\end{table}

Properties estimated by \textsc{pycprops} for the synthetic sources were compared to the known values used to inject the sources, over ranges similar to the range of properties estimated for clouds found in the original \gls{co} cube.
The values from \textsc{pycprops} are those for which an attempt to remove the effects of finite sensitivity and resolution have been applied (see section~3.4.2 of \citealt{Ros2021} for details of how these corrections are done), and we compare with the input synthetic cloud properties before they are adjusted for the effects of resolution.
The ratio of total recovered luminosity to the input luminosity is about \num{0.7}, on average, with a standard deviation near \num{0.2}.
When considering found clouds with the largest fraction of the luminosity, the ratio of estimated radii relative to input radii also averages around \num{0.7} with a standard deviation of about \num{0.2}.
The same comparison for velocity dispersions shows a ratio of about \num{0.8}, on average, with a standard deviation around \num{0.3}.
These biases and spreads are consistent with those reported by \citet{Ros2021} for the \gls{phangs} data.

\section{Results} \label{sec:results}
In this section we summarize the cloud properties measured in \ngc{} and compare to the cloud properties in \gls{phangs} galaxies.
Samples only include clouds that could be both spatially and spectrally deconvolved from the resolution of the observations (performed internally by \textsc{pycprops}).

\subsection{Two-dimensional model of clouds} \label{sec:2d_distributions}
\begin{figure*}
    \centering
    \includegraphics{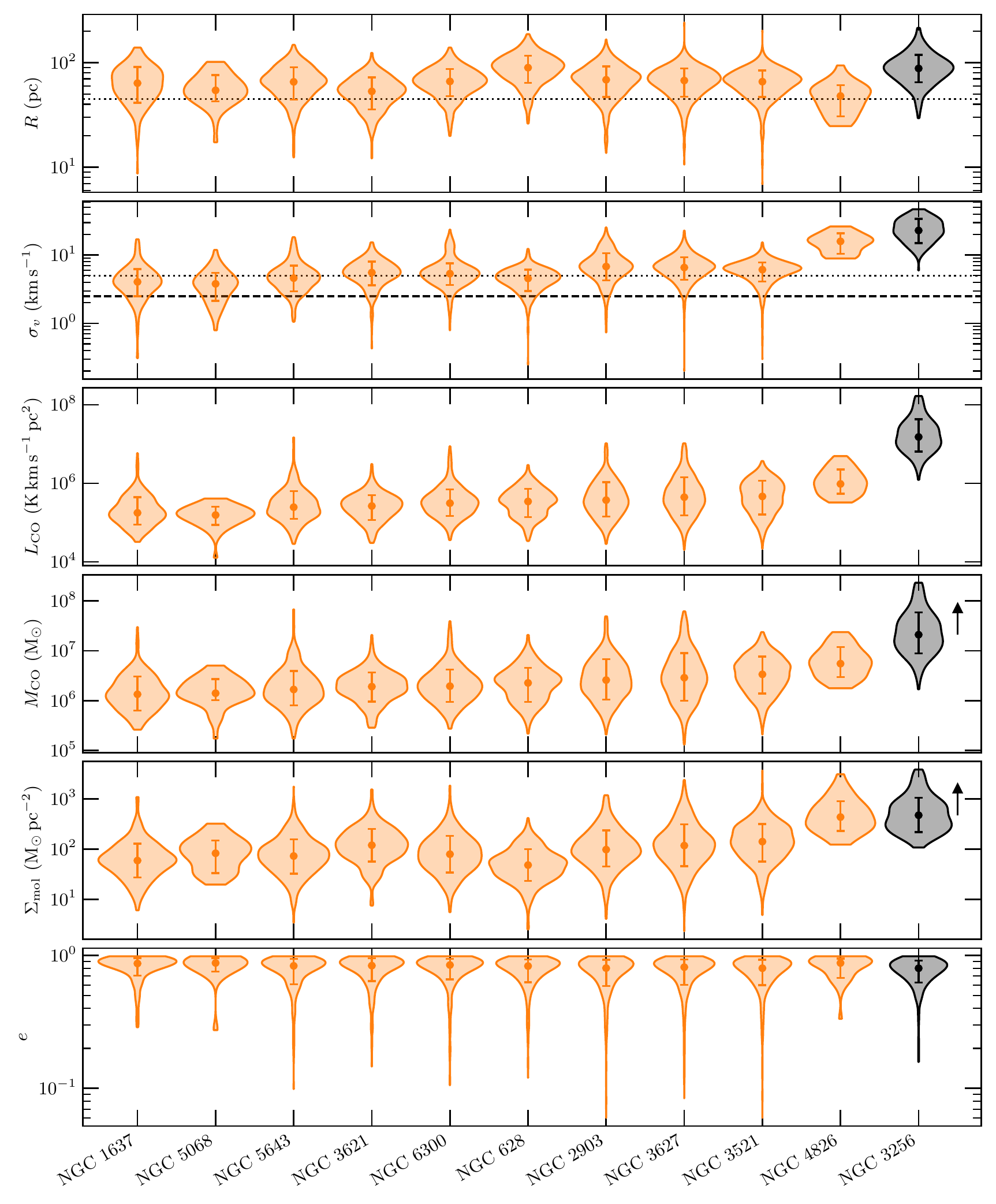}
    \caption{
        Resolved cloud-property distributions for \protect\ngc{} and the \gls{phangs} sample.
        Circles show medians and errorbars show \nth{16} to \nth{84} percentile ranges.
        Curves and shaded regions show Gaussian \glspl{kde}, with uniform weights for all clouds.
        The widths are normalized for each parameter to show relative cloud fractions per dex between galaxies (i.e. the area of the shaded regions is the same for each galaxy).
        Galaxies are sorted from low (left) to high (right) median cloud mass.
        The beam \gls{hwhm} of \SI{45}{\parsec} is shown with the dotted line in the $R$ panel.
        Dotted and dashed lines in the $\sigma_{v}$ panel show channel widths of \SI{5}{\kilo\metre\per\second} for \protect\ngc{} and \SI{2.5}{\kilo\metre\per\second} for \gls{phangs}, respectively.
        Masses and mass surface densities in \protect\ngc{} are estimated with a \gls{u/lirg} $\alpha_{\mathrm{CO}}$ of \SI{1.38}{\solarmass\per\square\parsec}(\si{\kelvin\kilo\metre\per\second})$^{-1}$ (others with the Milky-Way value).
        Multiplying the masses and mass surface densities in \protect\ngc{} by \num{\sim 4.5} would switch those values to using the Milky Way conversion factor, with these shifts shown by the arrows.
    }
    \label{fig:2d_kdes}
\end{figure*}

Figure~\ref{fig:2d_kdes} compares the distributions of cloud properties between \ngc{} and each \gls{phangs} galaxy, estimated directly from the \textsc{pycprops} measurements\footnote{
    The electronic table of clouds found by \citet{Ros2021} was retrieved from the journal website on \num{2021} July \num{31}.
}.
Note that all Gaussian \gls{kde} bandwidths presented here were automatically calculated using the \textsc{scipy} implementation of Scott's Rule \citep{Sco1992}, and that uniform weights were used for all clouds.
Most properties are significantly larger in \ngc{} than most or all of the clouds identified by \citet{Ros2021} (velocity dispersion, luminosity, \gls{co}-estimated mass, and mass surface density).
Smaller differences are present for the distributions of on-sky radii, and no significant differences appear in the distributions of estimated cloud eccentricities.
Medians and inner \nth{68} percentiles are shown as circles and errorbars, respectively, but the means are very similar to the medians for all galaxies and properties.
It is worth noting that the uncertainties on the means would appear smaller than the circles such that the mean for nearly every galaxy is significantly different from every other galaxy.

We used a single \gls{u/lirg} type conversion factor $\alpha_{\mathrm{CO(2-1)}} = \SI{1.38}{\solarmass\per\square\parsec}(\si{\kelvin\kilo\metre\per\second})^{-1}$ for \ngc{} \citep[see][for a description of how we decided this was appropriate]{Bru2021}.
For \gls{phangs}, a metallicity-dependent conversion factor was used based on radial metallicity gradients from \citet{San2014, San2019} and described in detail by \citet{Sun2020a}.
All conversion factors for the \gls{phangs} clouds were larger than the value we used for \ngc{}, with a median of \SI{\approx 6.75}{\solarmass\per\square\parsec}(\si{\kelvin\kilo\metre\per\second})$^{-1}$, such that all \gls{phangs} luminosities were scaled up by a larger factor to convert to masses than \ngc{}.
Switching our choice of conversion factor for \ngc{} to that of the Milky Way would result in the masses increasing by a factor of \num{\sim 4.5}, with the horizontal arrow in Figure~\ref{fig:2d_kdes} showing this change.

The most important comparisons are between the directly-observable properties (cloud size, velocity dispersion, peak brightness temperature, and luminosity), since the other properties are each derived from the former.
However, given the galaxy-specific scatter in the relations between these cloud properties (some of which are presented in Figures~\ref{fig:vd_vs_3d_radius} through \ref{fig:mvir_vs_mco}) we also show properties like mass surface density and eccentricity that are derived through combinations of the observables.
Finally, due to the lack of published peak brightness temperatures for the \gls{phangs} clouds, we cannot present a comparison of this property (though we do include it for \ngc{} in Table~\ref{tab:cloud_properties_1} and Figure~\ref{fig:pixel_cloud_kdes}).

There is substantial overlap between the two-dimensional radii estimated for clouds in \ngc{} and the \gls{phangs} galaxies, but the distribution for \ngc{} extends to larger radii than most \gls{phangs} galaxies.
The cloud-radius distribution in \ngc{} is very similar to that in \ngc{628}, both of which exhibit the largest clouds shown here.
The sizes of clouds in \ngc{} are also noteworthy because nearly all radii are significantly larger than the beam \gls{hwhm}, shown as the dotted line in Figure~\ref{fig:2d_kdes}.
Source-finding algorithms built on emission segmentation, such as \textsc{pycprops}, typically identify peaks of emission near the size of the beam \citep{Pin2009, Hug2013, Ler2016, Rei2010, Ros2021}, which is clearly apparent for many of the clouds in the \gls{phangs} galaxies.

Given the matched resolution to the \gls{phangs} cloud-finding analysis, identical source-finding algorithm, and similar treatment of signal and noise portions of our cube it appears the difference in the sizes of molecular gas structures between \ngc{} and most \gls{phangs} galaxies is real.
The difference in noise between our observations of \ngc{} and those from \gls{phangs} could result in differences in the measured cloud sizes.
However, noise levels have been shown to be anti-correlated with measured cloud size for clouds estimated to be the size of the beam or larger \citep{Ros2006}, which \num{96} per cent of our resolved clouds from \ngc{} are.
Bias related to higher noise would be making the clouds from \ngc{} appear smaller than they actually are so that the difference from the \gls{phangs} galaxies is more likely to be larger than estimated here rather than smaller.

Nearly all clouds in \ngc{} have velocity dispersions well above our spectral resolution of \SI{\approx 5}{\kilo\metre\per\second} (no clouds have velocity dispersions less than the channel width).
The same is true for most galaxies from \gls{phangs}, except for \ngc{5068} where \num{\approx 27} per cent of resolved clouds are estimated to have dispersions below their spectral resolution of \SI{2.5}{\kilo\metre\per\second}.
The majority of clouds from \ngc{} have velocity dispersions significantly larger than clouds from \gls{phangs}.
However, the upper half of the distribution from \ngc{4826} overlaps with the lower half of the distribution from \ngc{}.

Both the noise level and channel width are complicating differences between the velocity-dispersion measurements of \ngc{} and \gls{phangs}.
\citet{Ros2006} showed that velocity dispersions are overestimated in the presence of finite-width channelisation, by about \num{15} per cent for high-\gls{s/n} clouds when the true velocity dispersions are equal to the channel width
This bias rapidly increases for dispersions smaller than the channel width (see their figure~\num{1}).
They also explored varying the noise level relative to the cloud peak intensity, showing that overestimating the velocity dispersion was worse at lower \gls{s/n}.
For example, the same \num{15} per cent bias occurred for clouds with true dispersions about twice the channel width at a \gls{s/n} of about \num{5}.

If we assume the measured velocity dispersions from the \gls{phangs} galaxies are their true dispersions and use the estimates of velocity-dispersion bias from \citet{Ros2006} for clouds with \gls{s/n} of only \num{5}, then we estimate the channels would have to be about \SI{15}{\kilo\metre\per\second} wide to shift the median dispersions to be roughly the same as measured in \ngc{}.
Including clouds with a range of \gls{s/n} values above five would require even wider channels (the \gls{phangs} \gls{s/n} values were not published but we have included them for \ngc{} in Tables~\ref{tab:cloud_properties_1}).
Also, the distributions are extremely narrow, making the \gls{phangs} distributions look nothing like the wide distribution measured in \ngc{}.
While the differences between noise and channel width do not appear to be the obvious direct drivers of the differences in cloud size and velocity dispersion, these comparisons would be best explored by altering the \gls{phangs} cubes to have the same noise and channel widths as our observations of \ngc{} and performing cloud finding again.
An exploration of this type will be presented in a following paper on cloud finding in the Antennae galaxy merger.

We performed two-sample Anderson-Darling tests \citep{Sch1987} between \ngc{} and each \gls{phangs} galaxy for each of the cloud properties in Figure~\ref{fig:2d_kdes}.
The null hypothesis here is that both the sample from \ngc{} and the \gls{phangs} galaxy come from the same distribution.
The radius distributions between \ngc{} and \ngc{628} appear to originate from the same underlying distribution.
Eccentricity distributions from \ngc{628}, \numlist{2903; 3521; 3627} appear consistent with coming from the same underlying distribution as \ngc{}.
Lastly, the mass surface density distributions from \ngc{} and \ngc{4826}  also appear to come from the same parent distribution.
For all other combinations of galaxies and cloud properties, the null hypothesis is rejected at the five per cent level.
These properties for all clouds in \ngc{} (including unresolved clouds) are reported in Tables~\ref{tab:cloud_properties_1} and \ref{tab:cloud_properties_2}, and physical implications of these results are discussed in Section~\ref{sec:discussion}.
We also performed the same Anderson-Darling tests between each pair of \gls{phangs} galaxies and found that for most properties the samples appear to come from different underlying distributions.
There are some instances where the null hypothesis could not be rejected (most in eccentricity followed by radius), but there is no obvious systematic trend of which galaxy pairs are indistinguishable.
While the differences between the underlying distributions of \ngc{} and the \gls{phangs} galaxies are significant, the dissimilarity also between the \gls{phangs} galaxies means we cannot yet attribute the differences in \ngc{} solely to it being a merger.

\subsection{Cloud orientations}
\begin{figure*}
    \centering
    \includegraphics{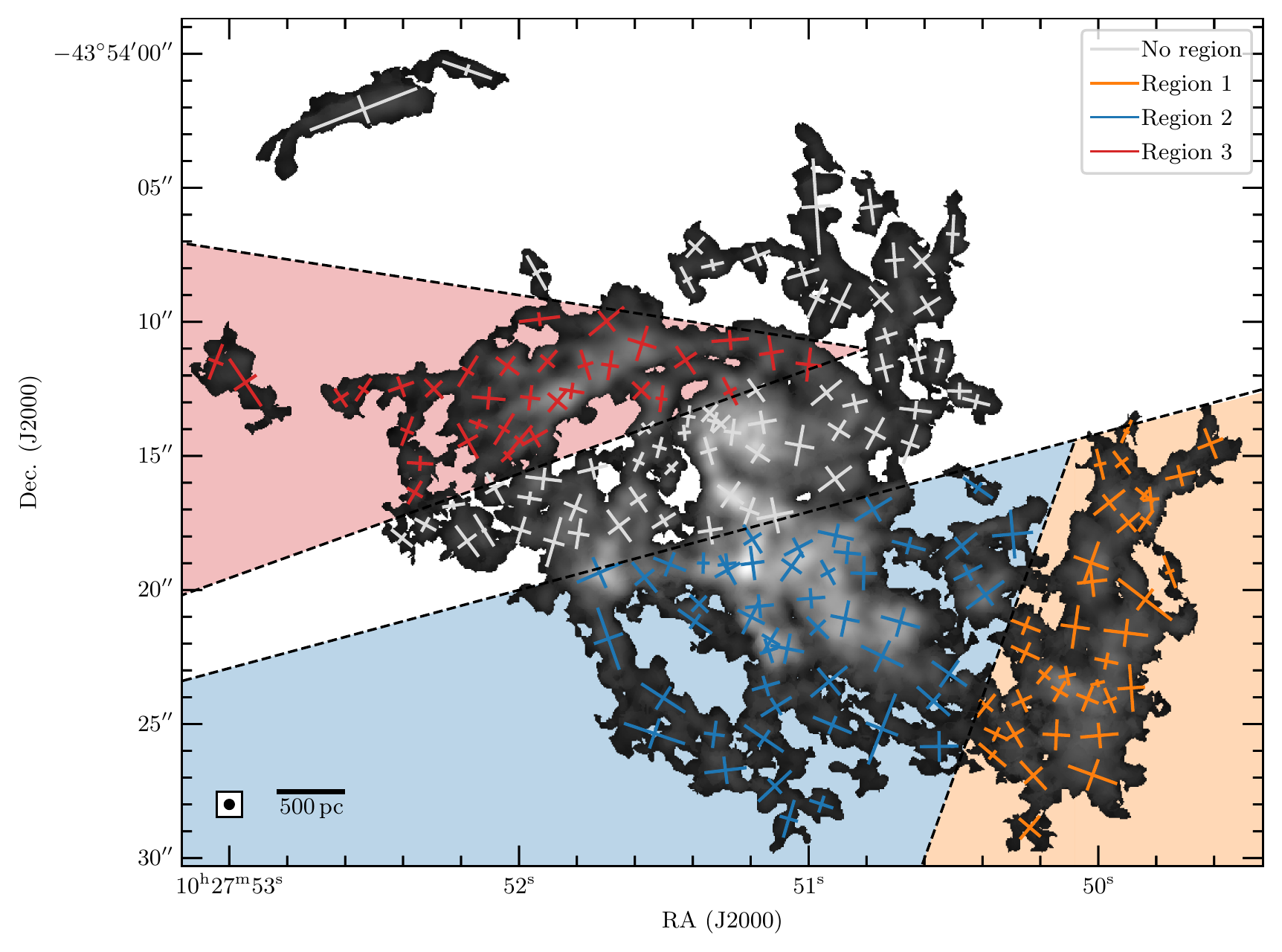}
    \caption{
        Integrated-intensity map of \protect\ngc{} calculated from the \gls{co} (\num{2}--\num{1}) cube on which the cloud finding was performed and with the same signal mask as used in cloud finding.
        Crosses mark the on-sky positions of all resolved clouds, and the length of the bars are equal to the estimated \gls{fwhm} major and minor axes of the clouds in the plane of the sky.
        Dashed black lines and colors mark the three regions selected by eye for separate tests of cloud alignment (light-gray crosses were not included in any of these by-eye regions).
        The circle in the bottom left shows the beam \gls{fwhm} of \SI{90}{\parsec}, and a scale bar is included to show \SI{500}{\parsec} at the distance of \protect\ngc{}.
    }
    \label{fig:position_angle_map}
\end{figure*}

Since the distribution of estimated cloud eccentricities is heavily skewed to elliptical clouds, we inspected the spatial distribution of the estimated \glspl{pa} to search for signs of cloud alignment.
Given the presence of some spiral structure around the northern nucleus and the edge-on orientation of the southern nucleus, it seemed possible that signatures of cloud alignment may exist due to shear or viewing angle.
Figure~\ref{fig:position_angle_map} shows the resolved-cloud positions with their \gls{fwhm} major and minor axes overlaid on an integrated intensity map of \ngc{} made from the same signal cube as the cloud finding was carried out on.

We checked the entire sample of resolved clouds across \ngc{} as well as three regions chosen by eye that appeared to potentially contain clouds more aligned than the entire sample.
Clouds in the regions chosen by eye are shown in Figure~\ref{fig:position_angle_map} in red, blue, and orange, and the region boundaries are shown with the dashed black lines and shaded regions.
\gls{ks} tests indicated that the distributions of \glspl{pa} in each of those samples were not distinguishable from a uniform distribution of angles from \ang{0} to \ang{180} (randomly oriented clouds).
Likewise, calculating the \gls{pa} differences between all pairs of clouds within those samples and using \gls{ks} tests to compare them to a uniform distribution of angles from \ang{0} to \ang{90} again could not rule out the differences being uniformly distributed.
Some small but statistically significant differences from uniform distributions were found when binning the \gls{pa} differences by the separations between the pairs of clouds.
However, no clear picture of cloud alignment is revealed since the separation bins were almost all $\gtrsim \SI{1}{\kilo\parsec}$.
It appears the disturbed morphology of this merger system has prevented any strong signal of cloud alignment being present in these data.
However, we hope these results will still be useful for future comparisons to other systems or theoretical studies.

\subsection{Cloud mass function}
\begin{figure}
    \centering
    \includegraphics{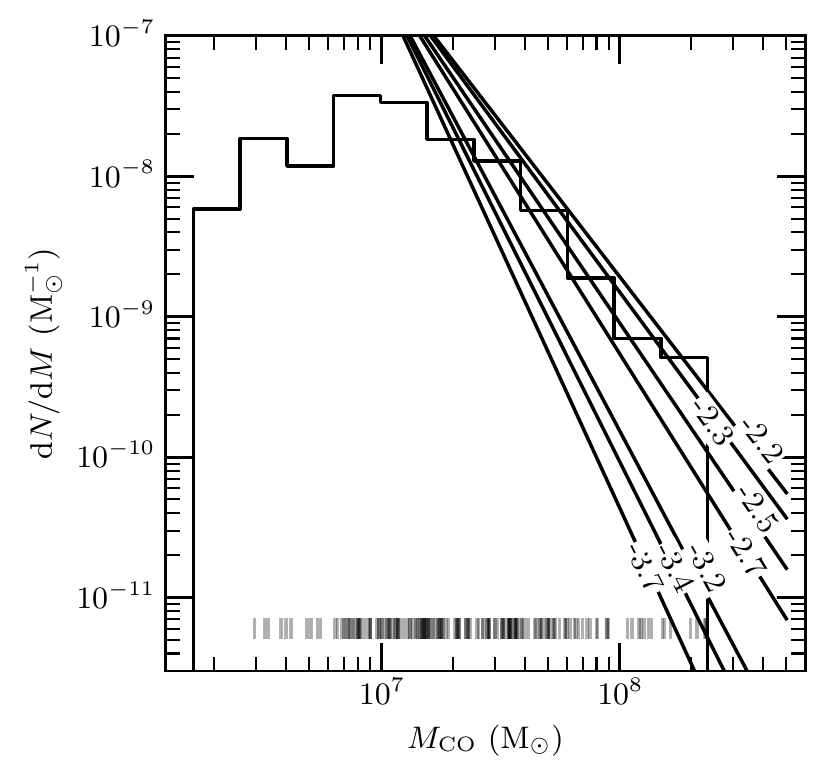}
    \caption{
        Differential mass function of clouds from \protect\ngc{} shown by the ``step-wise'' curve.
        Unbinned cloud masses are also shown as vertical lines near the bottom.
        Only resolved clouds are shown for consistency with the remaining figures and subsequent analysis.
        Straight lines indicate the range of pure power law indices from the \gls{phangs} galaxies fit by \protect\cite{Ros2021}.
    }
    \label{fig:differential_mass_function}
\end{figure}

\begin{figure}
    \centering
    \includegraphics{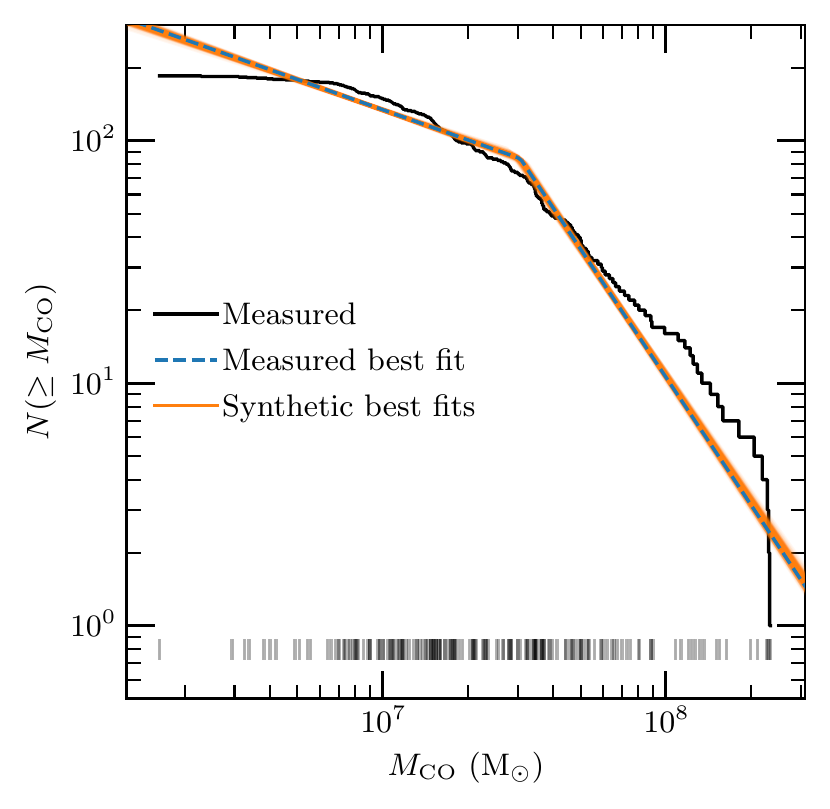}
    \caption{
        Cumulative mass function of resolved clouds from \protect\ngc{} shown by the ``step-wise'' curve.
        A one-dimensional view of the cloud masses is also shown as vertical lines near the bottom.
        The dashed blue curve shows the double power law best fit to the measured cloud masses.
        Orange curves show \num{200} of the \num{1e5} best fits to the synthetic cloud mass functions.
    }
    \label{fig:cumulative_mass_function}
\end{figure}

Figure~\ref{fig:differential_mass_function} presents the cloud mass function for \ngc{} in differential form; Figure~\ref{fig:cumulative_mass_function} presents the cumulative mass function.
The differential form shows the broad trend of a power law at high masses that turns over at low masses, while the cumulative form reveals some subtle features that are smoothed out in the differential binning.
To aid comparison with other studies we fit the cumulative mass function with a double power law of the form
\begin{equation} \label{eq:double_power_law}
    N(\geq M) =
    \begin{cases}
        A M_{\mathrm{break}}^{(\alpha_{\mathrm{high}} - \alpha_{\mathrm{low}})} M^{(\alpha_{\mathrm{low}} + 1)} & M < M_{\mathrm{break}} \\
        A M^{(\alpha_{\mathrm{high}} + 1)} & M \geq M_{\mathrm{break}}
    \end{cases}
\end{equation}
where $\alpha_{\mathrm{high}}$ is the high mass power law index, $\alpha_{\mathrm{low}}$ is the low mass power law index, $M_{\mathrm{break}}$ is the mass at which the power law index changes, and $A$ is a normalization constant \citep{Rei2006b}.
No bounds were applied to any of the free parameters during fitting.
We also chose to not apply any corrections to the numbers for completeness because the completeness is not well constrained by the mass alone and to facilitate comparison with the mass function shapes reported by \citet{Ros2021} who also did not modify the mass-function numbers.
Best-fitting parameters are shown in Table~\ref{tab:double_power_law_parameters}.
Following \citet[see their equation~\num{6} and preceding discussion]{Rei2006a}, the weight on each data point in the individual fits was set to $1 / N(\geq M_{\mathrm{CO}})^{2}$.

\begin{table}
    \centering
    \caption{
        Results of fitting the cumulative cloud mass function with a double power law.
        Best-fitting parameters come from fitting the measured mass function directly with weights of $1 / N(\geq M_{\mathrm{CO}})^{2}$.
        The parameter uncertainties were estimated by fitting \num{1e5} synthetic mass functions derived from the mass uncertainties and finding the \nth{5}, \nth{16}, \nth{84}, and \nth{95} percentiles of the resulting parameter distributions, $P_{5}$ through $P_{95}$.
        Estimated parameters correspond to those in Equation~\ref{eq:double_power_law}.
    }
    \begin{threeparttable} \label{tab:double_power_law_parameters}
        \begin{tabular}{
            @{}
            l
            S[table-format=+1.2]
            S[table-format=+1.2]
            S[table-format=+1.2]
            S[table-format=+1.2]
            S[table-format=+1.2]
            @{}
        }
            \toprule
            Parameter                                   & {Best-fitting value} & {$P_{5}$} & {$P_{16}$} & {$P_{84}$} & {$P_{95}$}  \\
            \midrule
            $\alpha_{\mathrm{low}}$                     & -1.41                & -1.43     & -1.42      & -1.40      & -1.39 \\
            $\alpha_{\mathrm{high}}$                    & -2.75                & -2.77     & -2.76      & -2.70      & -2.68 \\
            $M_{\mathrm{break}}$ (\SI{1e7}{\solarmass}) & 3.06                 & 2.88      & 2.95       & 3.11       & 3.17 \\
            $A / 10^{15}$                               & 1.1                  & 0.3       & 0.5        & 1.2        & 1.7 \\
            \bottomrule
        \end{tabular}
    \end{threeparttable}
\end{table}
% Table notes:
% - best-fitting values from single fit to measured masses
% - percentiles from the 10^5 synthetic mass function fits
% - choosing decimal places:
%    - alpha_low, alpha_high, and M_break need at least 3 sig figs (2 decimal places for each value shown to be distinguishable
%    - for A I prioritize showing the same decimal place across all values shown over the same number sig figs

Caution should be used when evaluating the quality of the fit by eye, as the logarithmic axes make it difficult to compare how much the fit deviates from the measurements at different masses.
Specifically, a visually-large vertical separation in the lower-right appears more significant than the upper-left, though the opposite is usually quantitatively true.
Also, the weighting of points is very strong towards the highest masses.
For example, the fit being above the measured mass function for the two most-massive clouds contributes the same fraction to the weighted residuals as the fit being below the measurements for the next \num{12} clouds that are above \SI{1e8}{\solarmass}.
The previous point is partially counteracted by the fact that the density of clouds along the $x$ axis is not uniform (or log uniform; see how the vertical lines at the bottom of Figure~\ref{fig:differential_mass_function} are clustered towards the centre of the figure).
This concentration of clouds between about \SIrange{1e7}{1e8}{\solarmass} means that the weights increase more rapidly per dex over that range than they would for a uniform (or log-uniform) distribution of cloud masses.
All of this together results in the fitting procedure trying very hard to match the high-mass end, and the extra flexibility around the break mass leading it to landing where the cloud masses are concentrated.

To incorporate the uncertainty on the mass of each cloud in the uncertainties on the best-fitting parameters we used a similar Monte Carlo approach to \citet{Rei2006a}.
We started by generating a new synthetic cloud-mass sample by drawing a random deviate from normal distributions centred at each measured cloud mass and with standard deviations equal to each mass uncertainty.
With a new sample of \num{185} masses we recalculated $N(\geq M_{\mathrm{CO}})$ and fit this synthetic mass function in the same way as the measured masses.
We carried this out for \num{1e5} synthetic mass functions and report the inner \numlist{68; 95} per cent of those best-fitting parameters in Table~\ref{tab:double_power_law_parameters}.

The uncertainties estimated in this way are likely an underestimate since they are only probing the parameter variations caused by our uncertainties on the masses, with all else held constant.
For example, while our choice of initial guess for the fit has little effect on the best-fitting parameters, the initial guess did have some impact on the distributions of the parameters from the synthetic mass-function fitting.
The particular initial guess we have used throughout the analysis presented here resulted in symmetric and nearly Gaussian distributions of best-fitting parameters.
Further discussion of the mass function, including comparisons to the \gls{phangs} galaxies, is in Section~\ref{sec:mass_function_discussion}.

\subsection{Estimating three-dimensional cloud sizes} \label{sec:estimate_r_3d}
Additional properties of the clouds can be derived if the three-dimensional structure of the clouds can be estimated.
\citet{Ros2021} estimate the three-dimensional cloud radius from the geometric mean in each dimension and with an adjustment made for clouds that exceed the characteristic molecular gas disc thickness.
They assume a molecular disc \gls{fwhm} of \SI{100}{\parsec} for all galaxies in their sample.
With that disc thickness they estimate the three-dimensional cloud radius, $R_{\mathrm{3D}}$, to be either the two-dimensional radius, $R$, if $R \leq \mathrm{FWHM} / 2$ or $\sqrt[3]{R^{2} \mathrm{FWHM} / 2}$ otherwise.
We calculate $R_{\mathrm{3D}}$ for the clouds in \ngc{} in the same way but we use a different molecular disc \gls{fwhm} that we estimate empirically for \ngc{} specifically.
We take the median of the scale heights for \ngc{} from \citet{Wil2019}, $H_{\mathrm{W}}$, and calculate the median molecular disc $\mathrm{FWHM} = 2 \sqrt{\ln{2}} H_{\mathrm{W}} \approx \SI{280}{\parsec}$.

\begin{figure*}
    \centering
    \includegraphics{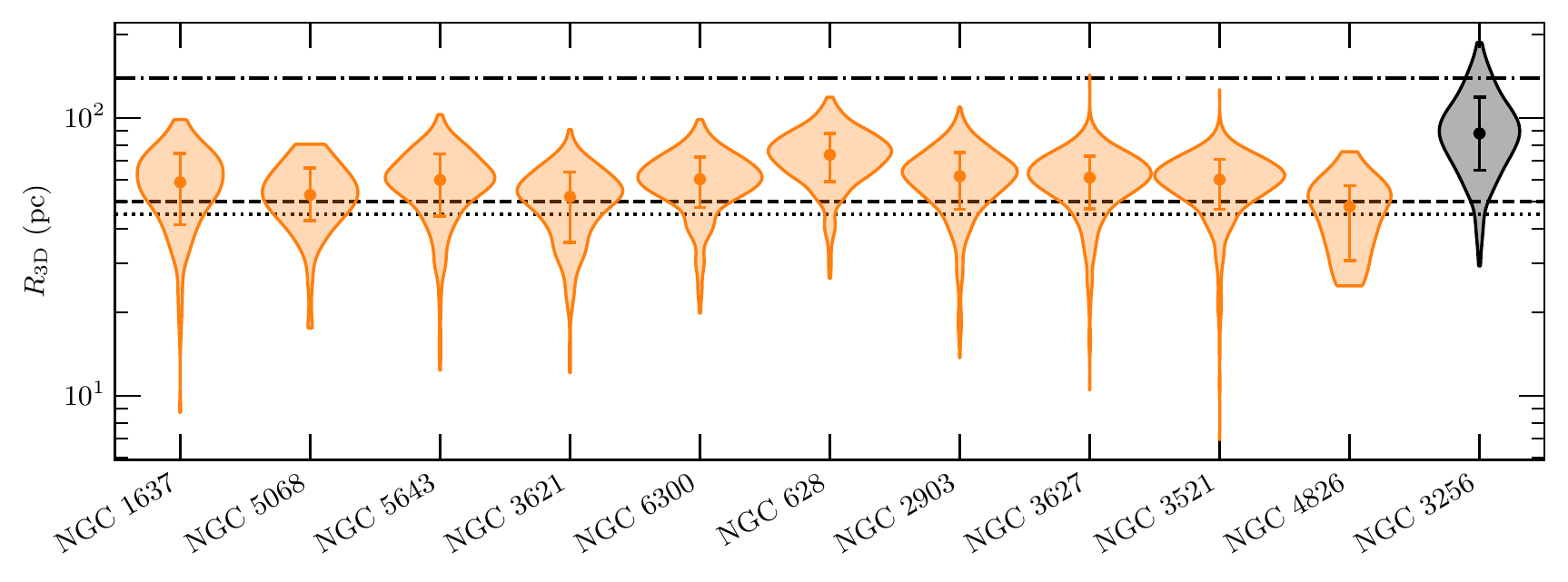}
    \caption{
        Same as Figure~\ref{fig:2d_kdes} but for estimated three-dimensional radii.
        Horizontal lines are, from bottom to top, the beam \gls{hwhm}, molecular disc \gls{hwhm} of \SI{50}{\parsec} used for \gls{phangs} galaxies, and disc \gls{hwhm} of \SI{140}{\parsec} for \protect\ngc{}.
    }
    \label{fig:3d_radii}
\end{figure*}

We calculate the three-dimensional radii for all of our clouds, shown in comparison with the \gls{phangs} clouds in Figure~\ref{fig:3d_radii}.
Horizontal lines, from bottom to top, show the beam \gls{hwhm}, the \gls{phangs} molecular disc \gls{hwhm} of \SI{50}{\parsec}, and the disc \gls{hwhm} for \ngc{} of \SI{\approx 140}{\parsec}.
Since the piece-wise form of $R_{\mathrm{3D}}$ acts most strongly on the clouds with the largest $R$, and all of the \gls{phangs} clouds are limited by the same disc thickness, the difference between the \ngc{} and \gls{phangs} distributions is more pronounced in $R_{\mathrm{3D}}$ than $R$.
Only about seven per cent of the resolved clouds in \ngc{} have $R \neq R_{\mathrm{3D}}$ while this is true for \num{78} per cent of resolved \gls{phangs} clouds.

\subsection{Three-dimensional model of clouds}
\begin{figure*}
    \centering
    \includegraphics{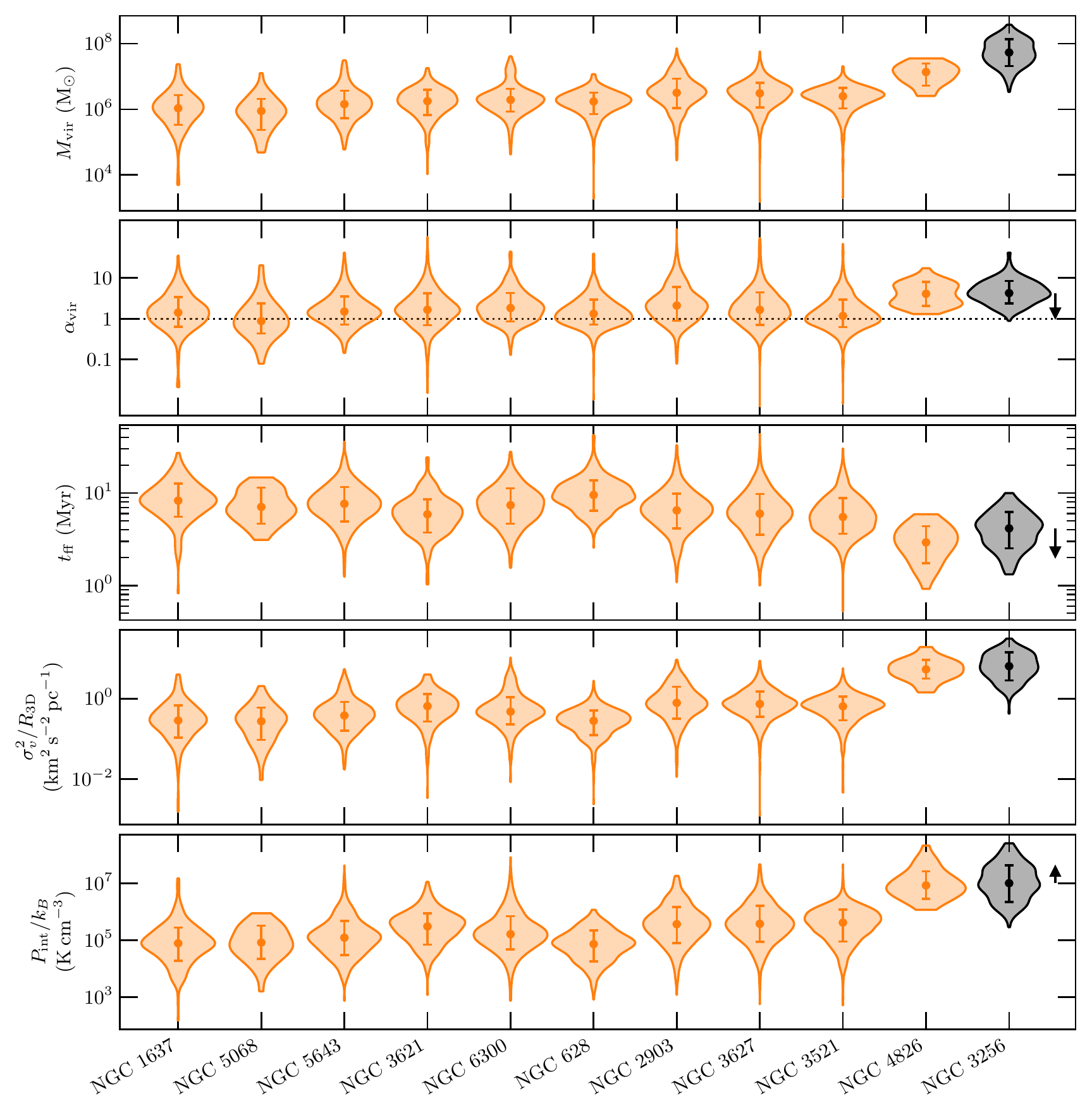}
    \caption{
        Same as Figure~\ref{fig:2d_kdes} but for properties derived using the three-dimensional radius estimates.
        The horizontal dotted line at $\alpha_{\mathrm{vir}} = 1$ indicates gravitational virial equilibrium.
        Calculations of the virial parameter, free-fall time, and internal pressures all depend on the choice of conversion factor.
        A \gls{u/lirg} conversion factor of \SI{1.38}{\solarmass\per\square\parsec}(\si{\kelvin\kilo\metre\per\second})$^{-1}$ is used for \protect\ngc{}.
        Horizontal arrows again show how the distributions from \protect\ngc{} that depend on the conversion factor would change if we switched to the Milky Way value.
        Quantitatively, to change to a Milky Way conversion factor would require dividing the virial parameters by \num{\sim 4.5}, dividing the free-fall times by \num{\sim 2}, and multiplying the pressures by \num{\sim 4.5}.
    }
    \label{fig:3d_kdes}
\end{figure*}

With an estimate for the three-dimensional radius of the \ngc{} clouds we calculated several additional properties (in combination with the directly-observable properties from Section~\ref{sec:2d_distributions}) and compare the distributions from \ngc{} to \gls{phangs} in Figure~\ref{fig:3d_kdes}.
Again, \ngc{} clouds appear at or above the upper limits of clouds found by \citet{Ros2021} for most of these quantities (virial mass, size-linewidth coefficient, and internal pressure).
However, the virial parameters for clouds in \ngc{} are similar to the \gls{phangs} galaxies.
Also, clouds in \ngc{} have some of the shortest estimated free-fall times, with only \ngc{4826} exhibiting shorter free-fall times.

We again performed two-sample Anderson-Darling tests for the properties presented in this section to test if the samples from \ngc{} and each \gls{phangs} galaxy appear to be drawn from the same underlying distribution.
The null hypothesis that the samples come from the same distribution was rejected at the \num{5} per cent level for all galaxies and properties except the virial parameters, size-linewidth coefficients, and internal pressures from \ngc{4826}.
As with the two-dimensional cloud properties however, the same tests between \gls{phangs} galaxies also indicated that most samples come from different underlying distributions.
These properties for all clouds in \ngc{} (including unresolved clouds) are reported in Tables~\ref{tab:cloud_properties_1} and \ref{tab:cloud_properties_2}, and discussion of possible physical interpretations of these results is in Section~\ref{sec:discussion}.

%%%%%%%%%%%%%%%%%%%%%%%%%%%%%%%%%%%%%%%%%%%%
% start of output from make_cloud_table.py %
%%%%%%%%%%%%%%%%%%%%%%%%%%%%%%%%%%%%%%%%%%%%

\begin{landscape}

\begin{table}
\centering
\caption{Example of cloud-property table showing the first ten rows and first \num{17} columns for the clouds found in \protect\ngc{} with \textsc{pycprops}.
The full version is available online in the Supporting Information.}
\begin{threeparttable}
\label{tab:cloud_properties_1}\begin{tabular}
{@{}S[table-format=2.0]S[table-format=1.1]S[table-format=2.1]S[table-format=5.0]S[table-format=3.4]S[table-format=+2.4]S[table-format=4.0]S[table-format=2.1]S[table-format=1.1]S[table-format=3.0]S[table-format=2.0]S[table-format=2.0]S[table-format=1.0]S[table-format=3.0]S[table-format=1.2]S[table-format=3.0]S[table-format=3.0]@{}}
\toprule
{Number} & {$T_{\mathrm{B,max}}$} & {$(S/N)_{\mathrm{max}}$\tnote{a}} & {$N_{\mathrm{pix}}$} & {R.A.} & {Dec.} & {$v$} & {$\sigma_{v}$} & {$\delta \sigma_{v}$\tnote{b}} & {$\sigma_{\mathrm{maj,d}}$} & {$\delta \sigma_{\mathrm{maj,d}}$\tnote{b}} & {$\sigma_{\mathrm{min,d}}$} & {$\delta \sigma_{\mathrm{min,d}}$\tnote{b}} & {P.A.} & {$e$} & {$R$} & {$R_{\mathrm{3D}}$\tnote{c}} \\
 & {(\si{\kelvin})} &  &  & {(\si{\degree}; J2000)} & {(\si{\degree}; J2000)} & {(\si{\kilo\metre\per\second})} & {(\si{\kilo\metre\per\second})} & {(\si{\kilo\metre\per\second})} & {(\si{\parsec})} & {(\si{\parsec})} & {(\si{\parsec})} & {(\si{\parsec})} & {(\si{\degree})} &  & {(\si{\parsec})} & {(\si{\parsec})} \\
\midrule
1 & 7.9 & 8.6 & 17744 & 156.9706 & -43.9034 & 2871 & 11.8 & 0.7 & 182 & 8 & 79 & 4 & 124 & 0.90 & 142 & 141 \\
2 & 5.2 & 5.8 & 8507 & 156.9710 & -43.9032 & 2872 & 16.2 & 0.9 & 117 & 9 & 43 & 4 & 69 & 0.93 & 83 & 83 \\
3 & 9.0 & 10.0 & 48931 & 156.9689 & -43.9006 & 2847 & 16.7 & 0.4 & 378 & 12 & 90 & 4 & 21 & 0.97 & 217 & 187 \\
4 & 6.8 & 7.9 & 5548 & 156.9674 & -43.9002 & 2825 & 5.8 & 0.5 & 166 & 14 & 28 & 3 & 161 & 0.99 & 81 & 81 \\
5 & 8.0 & 8.2 & 10597 & 156.9616 & -43.9016 & 2720 & 14.7 & 1.3 & 113 & 8 & 64 & 7 & 98 & 0.82 & 100 & 100 \\
6 & 11.7 & 12.5 & 18584 & 156.9624 & -43.9059 & 2740 & 33.4 & 1.8 & 92 & 4 & 61 & 4 & 135 & 0.75 & 88 & 88 \\
7 & 10.8 & 11.8 & 17945 & 156.9665 & -43.9036 & 2881 & 24.7 & 1.5 & 75 & 4 & 58 & 3 & 83 & 0.64 & 78 & 78 \\
8 & 7.4 & 8.5 & 6381 & 156.9584 & -43.9055 & 2660 & 15.9 & 1.9 & 93 & 9 & 92 & 11 & 97 & 0.16 & 109 & 109 \\
9 & 6.1 & 6.2 & 13402 & 156.9594 & -43.9067 & 2723 & 33.9 & 2.6 & 72 & 4 & 63 & 3 & 114 & 0.50 & 80 & 80 \\
10 & 8.7 & 9.5 & 6829 & 156.9643 & -43.9039 & 2788 & 23.2 & 1.5 & 43 & 4 & 34 & 3 & 96 & 0.62 & 45 & 45 \\
\bottomrule
\end{tabular}
\begin{tablenotes}
\item [a] Maximum signal-to-noise ratio over all signal and noise-cube pixels within each cloud.
\item [b] Uncertainties were estimated by \textsc{pycprops} through bootstrapping by resampling, with replacement, the pixels within each cloud \num{100} times, recalculating the cloud properties for each of those samples of pixels, and calculating the standard error on the mean of each property estimated from those samples. The relative uncertainties produced by \textsc{pycprops} were converted to absolute uncertainties for this table.
\item [c] Calculated from the $R$ column combined with the molecular disc scale height of \protect\ngc{} estimated by \protect\cite{Wil2019}. See Section~\ref{sec:estimate_r_3d} for details.
\end{tablenotes}
\end{threeparttable}
\end{table}

\begin{table}
\centering
\caption{Continuation of the example cloud-property table for the first ten rows and the last nine columns.
The cloud number is repeated as the left-most column for clarity.
The full version is available online in the Supporting Information.}
\begin{threeparttable}
\label{tab:cloud_properties_2}\begin{tabular}
{@{}S[table-format=2.0]S[table-format=2.1]S[table-format=1.1]S[table-format=2.1]S[table-format=3.0]S[table-format=2.1]S[table-format=2.1]S[table-format=2.0]S[table-format=2.1]S[table-format=1.2]@{}}
\toprule
{Number} & {$L_{\mathrm{CO}}$} & {$\delta L_{\mathrm{CO}}$\tnote{a}} & {$M_{\mathrm{CO}}$} & {$\Sigma_{\mathrm{mol}}$} & {$M_{\mathrm{vir}}$\tnote{c}} & {$\alpha_{\mathrm{vir}}$\tnote{c}} & {$t_{\mathrm{ff}}$\tnote{c}} & {$P_{\mathrm{int}} / k_{\mathrm{B}}$\tnote{c}} & {$P_{\mathrm{comp}}$\tnote{d}} \\
 & {(\SI{1e6}{\kelvin\kilo\metre\per\second\square\parsec})} & {(\SI{1e6}{\kelvin\kilo\metre\per\second\square\parsec})} & {(\SI{1e6}{\solarmass})} & {(\si{\solarmass\per\square\parsec})} & {(\SI{1e6}{\solarmass})} &  & {(\si{\mega\year})} & {(\SI{1e6}{\kelvin\per\cubic\centi\metre})} &  \\
\midrule
1 & 15.1 & 0.8 & 20.8 & 165 & 22.6 & 2.2 & 9 & 0.6 & 0.40 \\
2 & 6.0 & {\ldots\tnote{b}} & 8.2 & 190 & 25.2 & 6.1 & 6 & 2.2 & 0.07 \\
3 & 46.8 & 1.3 & 64.5 & 219 & 60.7 & 1.9 & 7 & 1.6 & 0.77 \\
4 & 5.4 & 0.5 & 7.4 & 180 & 3.2 & 0.9 & 6 & 0.3 & 0.60 \\
5 & 8.8 & 0.6 & 12.1 & 191 & 25.2 & 4.2 & 7 & 1.5 & 0.16 \\
6 & 24.9 & 1.3 & 34.3 & 705 & 114.0 & 6.6 & 3 & 32.8 & 0.64 \\
7 & 28.3 & 1.2 & 39.0 & 1032 & 55.2 & 2.8 & 3 & 29.9 & 0.94 \\
8 & 5.6 & 0.4 & 7.7 & 103 & 31.8 & 8.3 & 10 & 0.9 & 0.01 \\
9 & 11.6 & 0.5 & 16.0 & 404 & 106.0 & 13.2 & 4 & 21.4 & 0.11 \\
10 & 7.6 & 0.6 & 10.5 & 825 & 28.1 & 5.4 & 2 & 36.4 & 0.56 \\
\bottomrule
\end{tabular}
\begin{tablenotes}
\item [a] Uncertainties were calculated the same as described in Table~\ref{tab:cloud_properties_1}.
\item [b] Blank values indicate the value ``Not A Number'' was the result from \textsc{pycprops}.
\item [c] Calculated using the $R_{\mathrm{3D}}$ column.
\item [d] Completeness probability estimated with the logistic-function fit to the completeness results using the $M_{\mathrm{CO}}$, $\Sigma_{\mathrm{mol}}$, and $\alpha_{\mathrm{vir}}$ columns. See Section~\ref{sec:completeness} for details.
\end{tablenotes}
\end{threeparttable}
\end{table}

\end{landscape}

%%%%%%%%%%%%%%%%%%%%%%%%%%%%%%%%%%%%%%%%%%
% end of output from make_cloud_table.py %
%%%%%%%%%%%%%%%%%%%%%%%%%%%%%%%%%%%%%%%%%%

% Explanation (from meeting with Chris on September 1, 2021) for not including uncertainties on all quantities in cloud-properties tables
% - first, I think some of the uncertainties on the derived properties may be calculated incorrectly
%  - I have raised this with Erik on GitHub on August 4 but not heard back
%  - even if there are no errors, anything based on the virial mass or 3D quantities would have to have the uncertainties propagated and added to the table-generation code since pycprops does not handle those properties
% - second, we currently provide the uncertainties on the more "fundamental" measurements that cannot be derived from other columns/properties
%  - the rest of the uncertainties can be propagated by anyone with the table
%  - I/we would have to go through the propagation calculations and they would have to be added to the code that generates the paper and electronic tables

As the differences between the two-dimensional cloud properties in \ngc{} and \gls{phangs} galaxies appear real with the data available at present, we believe the differences in these properties estimated from the three-dimensional cloud model are also significant.
While it is unknown exactly how the emission would be segmented in \ngc{} if more sensitive observations were obtained, we know that the high peak brightness temperatures and total luminosity in clouds would remain.
It is clear that the molecular clouds in \ngc{} are extreme in their linear sizes, velocity dispersions, and masses which translates to likely harbouring large size-linewidth coefficients and internal pressures.
A clear picture on the dynamics of the gas is not possible with these observations alone, as the combination of a high \gls{sfr} with short free-fall times but also a higher but narrower distribution of virial parameters emphasize the need for a more complete accounting of confining gravitational components that include external forces \citet[e.g.][]{Sun2020a}.

\subsection{Correlations between cloud properties}
\begin{figure}
    \centering
    \includegraphics{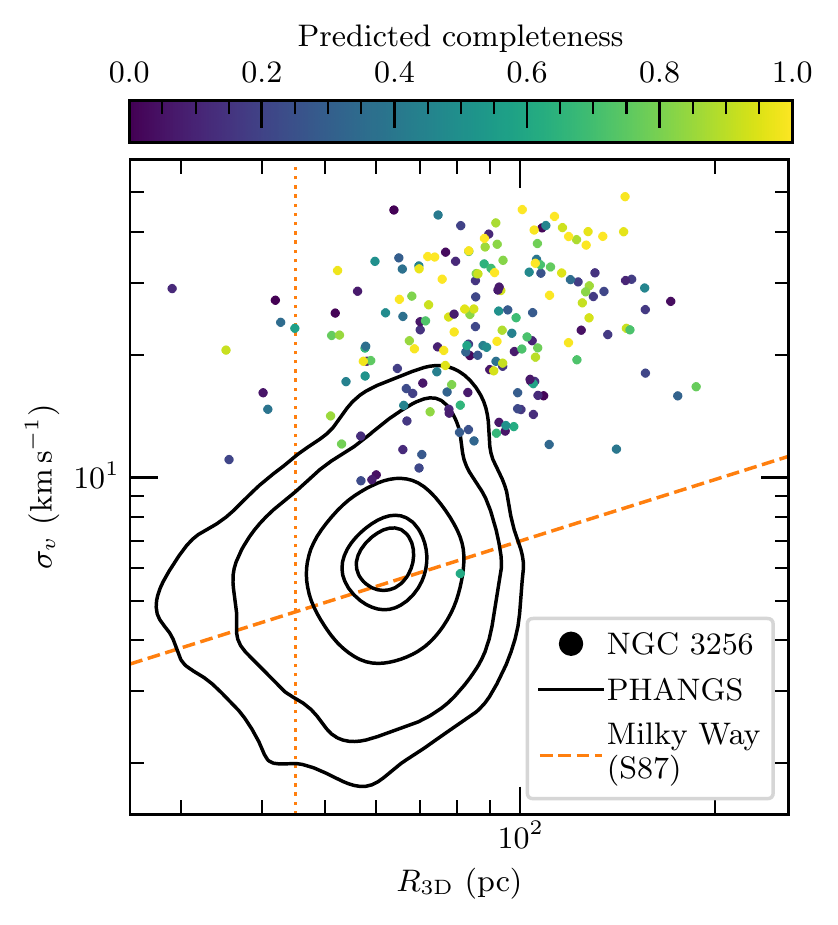}
    \caption{
        Velocity dispersion vs. three-dimensional radius.
        Each circle is a resolved cloud from \protect\ngc{}, and black contours enclose \numlist{90; 80; 50; 20; 10} per cent of the resolved \gls{phangs} clouds, determined by a Gaussian \gls{kde} with uniform weights for all clouds.
        Cloud circles from \protect\ngc{} are coloured by predicted completeness from the fit to the completeness test results and the cloud's estimated mass, surface density, and virial parameter.
        The orange-dashed line shows the fit to Milky Way clouds from \protect\cite{Sol1987}.
        The vertical-dotted line shows the beam \gls{hwhm} of \SI{45}{\parsec}.
    }
    \label{fig:vd_vs_3d_radius}
\end{figure}

\begin{figure}
    \centering
    \includegraphics{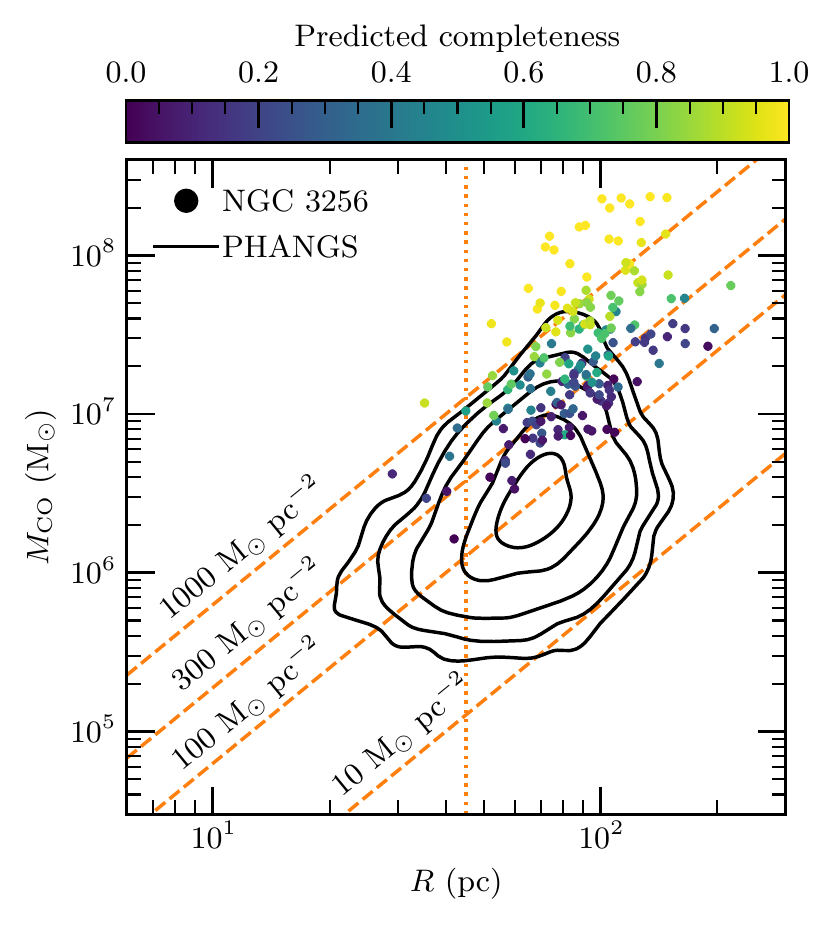}
    \caption{
        \gls{co}-estimated cloud mass vs. two-dimensional radius.
        Symbols are the same as in Figure~\ref{fig:vd_vs_3d_radius}, but contours here enclose \numlist{95; 90; 80; 50; 20} per cent of the \gls{phangs} clouds.
        The orange-dashed lines indicate constant mass surface densities.
        The vertical-dotted line shows the beam \gls{hwhm} of \SI{45}{\parsec}.
        A \gls{u/lirg} conversion factor of \SI{1.38}{\solarmass\per\square\parsec}(\si{\kelvin\kilo\metre\per\second})$^{-1}$ was used to calculate the \gls{co} masses in \protect\ngc{}; changing to the Milky Way conversion factor would shift those points up by \SI{\sim 0.6}{\dex}.
    }
    \label{fig:mco_vs_2d_radius}
\end{figure}

\begin{figure}
    \centering
    \includegraphics{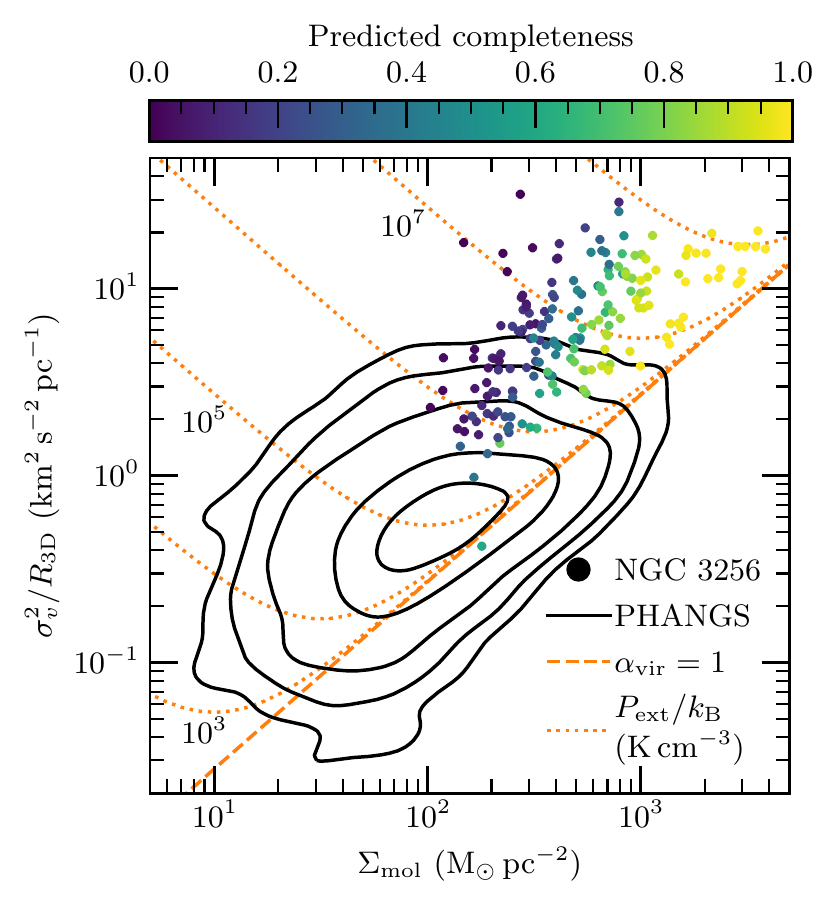}
    \caption{
        Size-linewidth coefficient vs. mass surface density.
        Symbols and contours are the same as in Figure~\ref{fig:mco_vs_2d_radius}.
        The orange-dashed line shows where clouds in virial equilibrium ($\alpha_{\mathrm{vir}} = 1$) would lie with no external pressure.
        Orange curving-dotted lines show where clouds would lie in virial equilibrium if instead they do experience external pressures, in units of \si{\kelvin\per\cubic\centi\metre}$\, k_{\mathrm{B}}^{-1}$.
        A \gls{u/lirg} conversion factor of \SI{1.38}{\solarmass\per\square\parsec}(\si{\kelvin\kilo\metre\per\second})$^{-1}$ was used to calculate the mass surface densities in \protect\ngc{}, and changing to the Milky Way conversion factor would shift those points to the right by \SI{\sim 0.6}{\dex}.
    }
    \label{fig:sl_coeff_vs_sd}
\end{figure}

\begin{figure}
    \centering
    \includegraphics{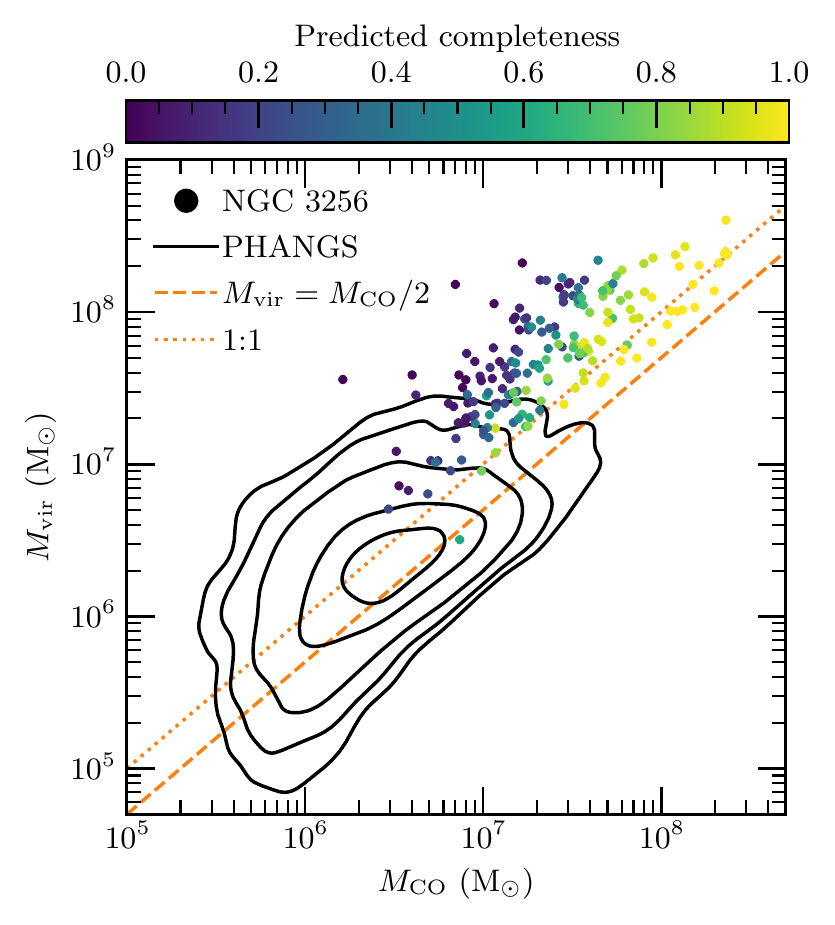}
    \caption{
        Virial mass vs. \gls{co}-estimated mass.
        Symbols and contours are the same as in Figure~\ref{fig:mco_vs_2d_radius}.
        The orange-dotted line shows where the virial mass is equal to the luminous mass and the dashed line shows where the virial mass is equal to half the luminous mass.
        The latter is expected for clouds in virial equilibrium when using two-dimensional Gaussian profiles for clouds, following \protect\cite{Ros2021}.
        A \gls{u/lirg} conversion factor of \SI{1.38}{\solarmass\per\square\parsec}(\si{\kelvin\kilo\metre\per\second})$^{-1}$ was used to calculate the \gls{co} masses in \protect\ngc{}, and changing to the Milky Way conversion factor would shift those points to the right by \SI{\sim 0.6}{\dex}.
    }
    \label{fig:mvir_vs_mco}
\end{figure}

In this section we plot several cloud properties against one another and compare any scaling relations found for clouds in \ngc{} to \gls{phangs}.
Generally, clouds found in \ngc{} appear separated in these parameter spaces from the \gls{phangs} population of clouds.
In Figures~\ref{fig:vd_vs_3d_radius} through \ref{fig:mvir_vs_mco}, individual clouds are shown as circles for \ngc{} and contours indicate the \gls{phangs} clouds, estimated from two-dimensional Gaussian \glspl{kde}.
Using the fit to the completeness results and the \textsc{pycprops}-estimated cloud mass, surface density, and virial parameter we predict the completeness for each cloud and show it as the colour of the circle.
The directions of trends in completeness are generally the same between \ngc{} and the \gls{phangs} galaxies.
Two additional versions of these scaling-relation plots are shown in Appendix~\ref{sec:more_scaling_relations}.
One version shows the \ngc{} clouds coloured by their distance from the nuclei, and the other highlights the \ngc{} clouds with the smallest radii or lowest velocity dispersions to show where these clouds appear in the different parameter spaces.
It is important to keep in mind how a change in the adopted conversion factor would change the results shown here, e.g. switching the \ngc{} points to the Milky Way conversion factor would move them \SI{\sim 0.6}{\dex} higher in \gls{co}-estimated mass or mass surface density.

Figure~\ref{fig:vd_vs_3d_radius} shows cloud velocity dispersions vs. estimated three-dimensional radii.
The orange-dashed line shows the fit to Milky Way clouds identified by \citet{Sol1987}.
Like the size-linewidth coefficients and internal pressures shown in Figure~\ref{fig:3d_kdes}, this figure shows how the velocity dispersions measured in clouds within \ngc{} exceed expectations from spiral galaxies, even given their larger sizes.
Given the limited dynamic range covered in this parameter space relative to the scatter of the data, it is not possible to say if the clouds from \ngc{} are forming a separate trend that is somehow offset and/or rotated relative to the Milky Way relation.
Even despite the large number of clouds identified by \gls{phangs}, their distribution of clouds does not show obvious evidence of a trend.
\citet{Ros2021} separate individual galaxies and then bin measurements by radius, which may indicate trends within galaxies that are offset between galaxies, but the significant scatter makes trends within galaxies hard to pin down.

\citet{Hug2013} caution that trends within the size-linewidth space should be carefully inspected for spatial and spectral resolution origins.
Given our matched spatial resolution with \gls{phangs} it appears real that the clouds in \ngc{} are slightly larger.
While our spectral resolutions are different by a factor of two, the difference between the median velocity dispersion and the channel width in \gls{phangs} is only about two while it is greater than a factor of four in \ngc{}.
The higher velocity dispersions in \ngc{} cannot be explained just by the wider channels.

Figure~\ref{fig:mco_vs_2d_radius} shows the \gls{co}-estimated masses vs. the two-dimensional estimated radii.
The orange-dashed lines show trends of constant mass surface density, with the \gls{phangs} distribution being centred on \SI{100}{\solarmass\per\square\parsec} and most of the clouds from \ngc{} centred on \SI{300}{\solarmass\per\square\parsec}.
The highest-mass clouds in \ngc{} actually scatter around mass surface densities of \SI{1000}{\solarmass\per\square\parsec}, but at the same radii there are also clouds down around \SI{300}{\solarmass\per\square\parsec} making the \ngc{} trend appear to fork at the largest radii.
The sparseness of clouds at these most extreme masses and sizes makes it difficult to tell if this split in the trend is real, especially given the strong dependence of the completeness at constant radius but varying mass.

The size-linewidth coefficients vs. mass surface densities are shown in Figure~\ref{fig:sl_coeff_vs_sd}.
The orange-dashed line indicates where clouds in virial equilibrium would lie without considering external pressure confinement, with both the \gls{phangs} and \ngc{} distributions clearly offset from that line (with the offset being greater for \ngc{}).
The curving dotted-orange lines show where clouds in virial equilibrium within a constant external pressure environment would appear \citep{Fie2011}.
Generally higher pressures are needed for the \ngc{} clouds to be kept in virial equilibrium.
The external pressure confined picture from Figure~\ref{fig:sl_coeff_vs_sd} is important to consider when interpreting the distributions of virial parameters in clouds from \ngc{} in Figure~\ref{fig:3d_kdes}, since external pressure is not included in the calculation of $\alpha_{\mathrm{vir}}$.
The clouds in \ngc{} may appear far from virial equilibrium, with far too much kinetic energy; however with sufficient confining pressure they could be held together and even made to collapse.

The relation between virial mass and \gls{co}-estimated mass is shown in Figure~\ref{fig:mvir_vs_mco}.
Both \gls{phangs} and \ngc{} clouds are shifted to higher virial masses than expected for virial equilibrium, but the significantly higher velocity dispersions in \ngc{} mean its clouds are shifted further.
Similar considerations around the presence of significant external pressure in \ngc{} described for the size-linewidth coefficients should be kept in mind when interpreting the virial masses.

\section{Discussion} \label{sec:discussion}
\subsection{Cloud properties that differ between NGC~\num{3256} and PHANGS-ALMA}
% - velocity dispersion
% - size-linewidth coefficient
In Figure~\ref{fig:2d_kdes}, the velocity dispersions of the clouds found in \ngc{} stand out as most dissimilar to the clouds found in the \gls{phangs} sample by \citet{Ros2021}.
Figure~\num{5} from \citet{Ros2021} shows clouds from the centres of galaxies have consistently the highest velocity dispersions, overlapping with the low end of the \ngc{} distribution.
In a pixel-based analysis, \citet{Bru2021} found that the centres of \gls{phangs} galaxies were most similar to the non-nuclear regions in \ngc{}.
The magnitude of the velocity dispersions measured across (most) of the \gls{phangs} sample are expected to originate primarily from star-formation driven turbulence \citep{She2012, Kru2018}.
However, to reach the velocity dispersions measured in \ngc{}, \citet{Kru2018} suggest gas flows into the centres of galaxies are also needed.
Interaction-induced torques drive the gas to flow towards the centres of merging systems \citep{Nog1988, Mih1996, Ion2004} which, in the model by \citet{Kru2018}, likely explains the enhanced velocity dispersions in \ngc{}.

Gas flows driven by interactions may also explain the high velocity dispersions in \ngc{4826}, since it is thought to have undergone a merger in the past that drove most of the \gls{ism} into the centre of the galaxy \citep{Bra1994}.
To a lesser extent this may also explain \ngc{3627} having the third highest median velocity dispersion in the \gls{phangs} sample due to its likely interaction with \ngc{3628} \citep{Rot1978, Soi2001}.

% velocity dispersion vs. size
It is interesting that most clouds in \ngc{} appear with significantly larger velocity dispersions and many with slightly larger radii, compared to \gls{phangs} clouds.
In a similar comparison at matched resolution (\SI{\sim 53}{\parsec} with \SI{5}{\kilo\metre\per\second} wide channels) and sensitivity, \citet{Hug2013} found that clouds in the interacting \ngc{5194} system have a distribution of radii very similar to \ngc{598} and the Large Magellanic Cloud but elevated velocity dispersions.
They observed that the regions of \ngc{5194} that harboured higher velocity dispersions were also where the \gls{sfr} was lower.
They argued that non-transient clouds are likely to be in a state that balances their internal turbulence with external pressure as well as gravity (satisfying $\sigma_{v} \propto P_{\mathrm{ext}}^{1/4} R ^{1/2}$).
They concluded that the elevated velocity dispersions at the same cloud sizes were driven more by differing external pressures between galaxies rather than different amounts of star-formation driven internal turbulence.
However, in \ngc{} we find a mix with clouds having both similar and larger radii to \gls{phangs} while almost always having larger velocity dispersions.
In \ngc{}, enhanced external pressures alone could not simultaneously increase the velocity dispersions and radii, so there are likely additional sources of turbulence in \ngc{} relative to \gls{phangs}.

If the external pressures were similar between \ngc{} and \gls{phangs} and Larson's size-linewidth relation held, then increased velocity dispersions would result in larger radii that lie along the dashed line in Figure~\ref{fig:vd_vs_3d_radius}.
The vertical offset of the clouds in \ngc{} from the dashed line to higher velocity dispersions may then indicate higher external pressures in \ngc{}.
Shown another way, clouds from \ngc{} are offset to higher internal pressures relative to \gls{phangs} in Figure~\ref{fig:3d_kdes}.
If the clouds are non-transient objects then the external pressures would have to be higher just to keep them bound, let alone collapse to form stars.

It is also interesting to qualitatively compare the integrated-intensity maps of \ngc{} \citep[figure~\num{1} from][]{Bru2021} and \ngc{4826} \citep[figure~\num{6} of the supplementary material from][]{Ros2021} because both galaxies exhibit smoother emission across the \gls{fov} compared to the much more clumpy appearance of the rest of the \gls{phangs} maps.
However, the \gls{fov} mapped in \ngc{4826} only reaches a radius of about \SI{1}{\kilo\parsec}, the same radius used to define the central regions of \gls{phangs} galaxies by \citet{Sun2018}.
Thus, the clouds from \ngc{4826} come solely from the central portion of the galaxy which \citet{Bru2021} found was the part of \gls{phangs} galaxies most similar to \ngc{}.

% - CO luminosity
% - CO-estimated mass
% - CO-estimated mass surface density
The \gls{co} luminosities of the clouds in \ngc{} are also significantly larger than those in the \gls{phangs} galaxies, even compared to \ngc{4826}.
\citet{Bru2021} found pixel-based peak brightness temperatures in \ngc{} reaching up to \SI{37}{\kelvin} (similar to those we find within clouds shown in Table~\ref{tab:pixel_cloud_percentiles} and Figure~\ref{fig:pixel_cloud_kdes}) compared to \SI{15}{\kelvin} in \gls{phangs} galaxies \citep{Sun2018, Sun2020b}.
Combined with the large velocity dispersions in \ngc{}, it is not surprising the \gls{co} luminosities are extreme as well \citep{Bol2013}.
From these luminosities, cloud masses are estimated to be significantly larger than in the \gls{phangs} sample, despite using a \gls{u/lirg} conversion factor that is between \numlist{3; 11} times smaller than the values used by \citet{Ros2021}.
While the largest masses may originate from clouds blended in space and velocity near the nuclei, \glspl{gmc} of masses \SI{\sim 1e8}{\solarmass} are likely necessary to produce star clusters with masses similar to globular-cluster progenitors \citep{How2017}.
\citet{Ada2020} measured the upper-mass cutoff to the cluster mass function in \ngc{} to be among the highest in local galaxies, which is consistent with this prediction of massive clouds resulting in the formation of massive star clusters.

Combining these mass estimates with the fairly similar cloud sizes seen in \ngc{} and \gls{phangs}, we see the molecular gas mass surface densities in \ngc{} exceed those measured in \gls{phangs} galaxies.
\citet{Kru2012} predicts a correlation between the mass surface density of molecular gas and the \gls{cfe}.
\citet{Ada2020} find general agreement between this theoretical prediction and their observations of nearby mergers as well as nearby spiral galaxies from the literature.
Specifically, \citet{Ada2020} estimate an upper limit on the \gls{cfe} in \ngc{} that is higher than most local galaxies.
If the surface density-\gls{cfe} correlation prediction by \citet{Kru2012} holds in general, then the extreme molecular gas mass surface densities in \ngc{} relative to the \gls{phangs} galaxies would imply its \gls{cfe} is high relative to the \gls{phangs} sample as well.

% - virial mass
% - virial parameter
Virial mass comparisons are very similar to those for velocity dispersion, given the strong dependence of virial mass on the velocity dispersion of the cloud.
Interestingly, the virial-mass distribution from \ngc{} is one of the widest shown in Figure~\ref{fig:3d_kdes}, potentially exhibiting blended double peaks.
However, when the virial masses are combined with the \gls{co} masses to estimate the virial parameter, the resulting distribution from \ngc{} is one of the narrowest.
This implies most of the clouds in \ngc{} have their mass closely tracking with their size and velocity dispersion so that a relatively narrow range of dynamical states are present, compared to some of the distributions of clouds found in the \gls{phangs} galaxies.

\ngc{4826} has virial masses more closely approaching \ngc{} than the other galaxies.
The two galaxies have a similar median virial parameter, but \ngc{4826} exhibits a different interplay between virial mass and virial parameter.
Although there is a hint of a multi-modal distribution of virial masses in \ngc{4826} with a peak clearly around \SI{1.5e7}{\solarmass} and a possible second peak near \SI{3e6}{\solarmass}, it is clearly multi-modal in virial parameter.
The lower virial-parameter peak in \ngc{4826} around \num{2} could indicate collapsing or marginally unbound gas while the second peak near \num{7} is either very unbound or requires a significant external-pressure contribution to remain bound.
Given the relatively broad distributions of the other virial-parameter distributions from the \gls{phangs} galaxies (which can be seen by comparing the widths of the violins in Figure~\ref{fig:3d_kdes}), perhaps we are seeing evidence that the \gls{ism} in these local spiral galaxies contains gas in a wider variety of dynamical states than the gas in \ngc{}.

% - internal turbulent pressure
The internal turbulent pressures in \ngc{} not only appear significantly higher than \gls{phangs} (except \ngc{4826}) but also exhibit the widest distribution with about three blended but distinct peaks.
If we assume most of the molecular gas is near pressure equilibrium, as \citet{Sun2020a} found with a subset of the \gls{phangs} sample, then this would imply there is also a wider range of external pressures present in \ngc{} than the nearby spiral galaxies.
For example, the violent rearrangement of gas through the merger process with significant mass inflow towards the nuclei could enhance external pressures.
At the same time, gas in the outskirts of the progenitors of \ngc{} would likely be less perturbed.
It is intriguing that \ngc{} has one of the narrowest distributions of virial parameter since the merger-driven rearrangement of the gas should also impact the state of the gas within clouds.
Processes seem to be conspiring to take the morphological mess that has been made of the gas and make its dynamics conform to a smaller range of states.
% In case someone wants some explanation of the virial parameter distribution being narrower in \ngc{}, here's some notes from my thoughts and Chris' comments.
% - Maybe I can come up with a better explanation based on the amount of energy injected by tidal shocks depending on the density of the region like described in section~\num{2.7} of \citet{Kru2012}
%  - I just couldn't figure out how to get the log distribution of virial parameters to become more narrow at the time.
% - From Chris: What if $\alpha_{\mathrm{vir}$ are enhanced in some parts of the \gls{phangs} discs because $\sigma_{v}$ broadened by rotation curve?
%  - And maybe in \ngc{} those clouds don't exist (sheared apart; or reduced shear)
% From Chris: Maybe narrow $\alpha_{\mathrm{vir}$ because all gas is self-gravitating, while in \gls{phangs} some ``clouds'' aren't (are transient)?

\subsection{Cloud properties that are similar between NGC~\num{3256} and PHANGS-ALMA}
% - radii
The cloud radii are one of the more consistent quantities between \ngc{} and \gls{phangs} galaxies.
Given the matched resolution of all the observations, and the tendency of algorithms like \textsc{pycprops} to identify structures near the resolution limit of the data, it is not too surprising how similar the distributions of radii between these galaxies are.
On the other hand, \ngc{} has the second largest median radius (second to \ngc{628}) and about \num{25} per cent of its clouds have radii exceeding most of the rest of the \gls{phangs} galaxies.
While \textsc{pycprops} attempts to remove observational effects in most of its estimates of cloud properties, our tests of the accuracy of the radius estimates (albeit without the effects of source blending; see Section~\ref{sec:completeness}) indicated it was systematically underestimating the true radii by about \num{30} per cent, meaning the true distribution of cloud radii may be shifted to even larger values.
\citet{Ros2021} do not find a systematic bias in the estimates of the radii in the \gls{phangs} data, so a similar shift to the true cloud radii is not expected for the \gls{phangs} distributions.
Finally, the higher noise level in our observations of \ngc{} would likely truncate the full spatial extent of clouds since \gls{s/n} cuts are used to select significant emission.
Better sensitivity would lead to more of the \gls{fov} being filled with significant emission, but it is ultimately difficult to predict how this would affect the radius distribution because \textsc{pycprops} would likely still be identifying sources near the beam size.

% - eccentricities
\citet{Dob2011} presented a connection between the aspect ratios of simulated clouds and their estimated degree of binding by self-gravity.
They found that their populations of simulated clouds with lower virial parameters appeared more spherical and regularly shaped than populations with elevated virial parameters in simulations with greater levels of stellar feedback.
Additionally, their simulations which best matched observed distributions of virial parameters in Galactic clouds also matched the observed distribution of Galactic cloud aspect ratios \citep[with the majority between \numrange{1.5}{2}; e.g.][]{Kod2006}.
\citet{Dob2011} gave an example of a long-lived and massive cloud which at first appeared fairly filamentary when its virial parameter was highest.
At later times, it appeared much more round and its virial parameter had dropped by a factor of about three.
In the observations presented here, we argue that the meaningful point to take away is that the molecular structures found in \ngc{} appear slightly larger than \gls{phangs} but the distributions of shapes are indistinguishable.
This may indicate that the enhanced velocity dispersions act to ``puff up'' the clouds in \ngc{} but on average their dynamical state is similar to clouds in nearby spiral galaxies.
It is important to note that at any instant an individual cloud's aspect ratio will not necessarily predict whether it will remain bound, but with a reasonable cloud sample size it may be possible to say if the size scale being probed is the primary size of objects that are bound.

% - free-fall time
\subsection{Free-fall times}
A noteworthy feature of the free-fall times in \ngc{} is that they are actually not as different as might be expected given the higher surface densities compared to the \gls{phangs} galaxies.
Since the free-fall time is proportional to the inverse of the volume density it appears the slightly larger cloud radii overcome some of the differences in mass so that the distributions of free-fall times overlap considerably.

\citet{Wil2019} estimated free-fall times ranging from \SIrange{2.5}{14}{\mega\year} in \ngc{} at \SI{512}{\parsec} resolution.
Despite our linear resolution being almost \num{6} times smaller than those observations (or \num{32} times smaller in area) we estimate free-fall times ranging from \SIrange{1}{10}{\mega\year}.
These similar free-fall times would imply similar average molecular gas volume densities between \SIlist{90; 512}{\parsec} scales.
Fairly constant gas properties across these size scales is also consistent with the minimal changes in pixel-based estimates of molecular gas surface density, velocity dispersion, and peak brightness temperature found by \citet{Bru2021} in \ngc{} at scales from \SIrange{55}{120}{\parsec}.
The interpretation that the molecular \gls{ism} in \ngc{} may be relatively smooth on the scales analysed by \citet{Bru2021} may also extend up to scales of about \SI{500}{\parsec}, since otherwise differing filling factors would result in measured gas densities changing with resolution.
For example, \citet{Sun2018} find more significant trends in these properties in their pixel-based analysis of \gls{phangs} galaxies at scales from \SIrange{45}{120}{\parsec}, indicative of the clumpy nature of the \gls{ism} in nearby spiral galaxies.

The efficiency per free-fall time is set by the ratio of the free-fall time to the gas depletion time ($\epsilon_{\mathrm{ff}} \equiv t_{\mathrm{ff}} / t_{\mathrm{dep}}$ where $t_{\mathrm{dep}} = \Sigma_{\mathrm{mol}} / \Sigma_{\mathrm{SFR}}$).
While the distribution of free-fall times in \ngc{} reaches much smaller values than most of the \gls{phangs} clouds from \citet{Ros2021}, about half the distribution overlaps with most of the \gls{phangs} free-fall times.
It seems that it is the difference in depletion times between \ngc{} and the \gls{phangs} galaxies that plays the dominant role in producing values of $\epsilon_{\mathrm{ff}}$ that are almost an order of magnitude larger in \ngc{} compared to spiral discs at \SI{500}{\parsec} scales \citep{Wil2019}.

The question of how the molecular-gas depletion times in \ngc{} compare at cloud scales to previous observations around \SI{500}{\parsec} remains open.
While the molecular mass surface densities measured here at \SI{90}{\parsec} resolution are comparable to those at \SI{512}{\parsec} resolution \citep{Wil2019}, measurements of the \gls{sfr} surface density at \SI{90}{\parsec} resolution are lacking.
In principle, continuum measurements from the observations presented here could give an estimate of the \glspl{sfr} in the regions of highest molecular gas surface density.
However, continuum measurements using solely these \SI{\sim 230}{\giga\hertz} observations will include significant contamination from dust emission.
The addition of observations at \SI{100}{\giga\hertz} would help extract just the free-free component for estimating the \gls{sfr}.

With direct estimates of the free-fall times from this work and depletion times from the continuum, it would be possible to estimate $\epsilon_{\mathrm{ff}}$ on cloud scales.
At \SI{90}{\parsec} resolution, it may be that the difference in $\epsilon_{\mathrm{ff}}$ between \glspl{u/lirg} and spirals is not as large as at \SI{\sim 500}{\parsec} resolution.
Unfortunately, a comparison of $\epsilon_{\mathrm{ff}}$ between spiral galaxies and starbursts on the scale of \glspl{gmc} is complicated by the stochastic nature of star formation at those physical scales.
However, it does raise the interesting possibility that the degree of stochasticity at a given scale may depend on the absolute level of the \gls{sfr} in the system.
For example, if the \gls{sfr} surface density were ten times higher in a particular galaxy, then about ten times the number of star-forming sites per unit area would be present.
In this situation, the area that would have to be averaged over to fully sample all star-formation stages to obtain an accurate estimate of $\epsilon_{\mathrm{ff}}$ could be about ten times smaller than in a galaxy with a lower \gls{sfr} surface density.
Imaging the \gls{co} and continuum observations at a range of resolutions and recalculating free-fall times, depletion times, and $\epsilon_{\mathrm{ff}}$ would allow us to explore the scatter in these quantities as a function of physical scale.

\subsection{Pixel-based vs. cloud-based emission decomposition}
\begin{figure}
    \centering
    \includegraphics{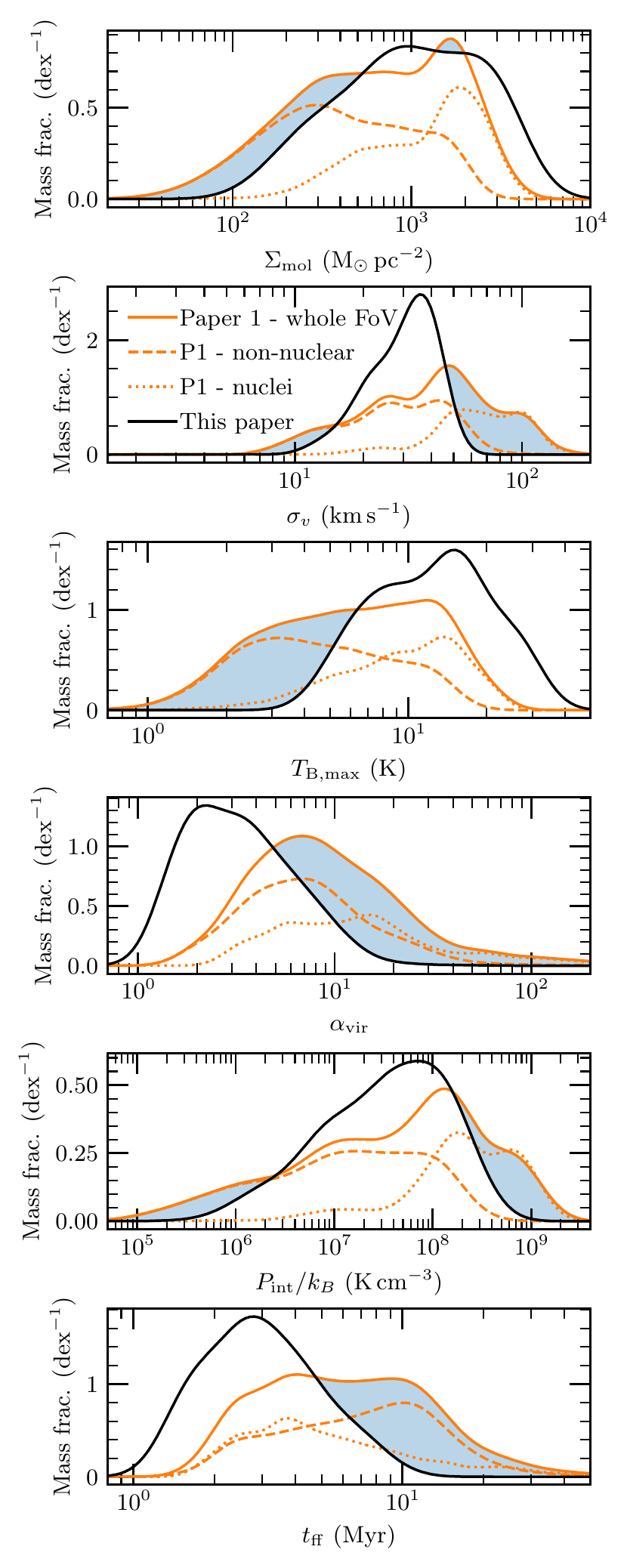}
    \caption{
        Mass-weighted Gaussian \glspl{kde} comparing molecular gas properties measured in clouds at \SI{90}{\parsec} resolution (black) to those measured from the pixel-based analysis from \protect\cite{Bru2021} at \SI{120}{\parsec} resolution (orange).
        Distributions from the pixel analysis are also broken into the non-nuclear (dashed) and nuclear (dotted) components.
        Blue-shaded regions show where the pixel distribution is above that from the clouds, to highlight where the distributions are most different.
        The same \gls{u/lirg} \gls{co}-to-H$_{2}$ conversion factor was used in both analyses.
    }
    \label{fig:pixel_cloud_kdes}
\end{figure}

Figure~\ref{fig:pixel_cloud_kdes} shows that there is general agreement between the cloud-finding results presented here and the pixel-based decomposition of these observations from \citet{Bru2021}\footnote{
    To avoid lines of sight with multiple spectral components, \citet{Bru2021} excluded pixels west of an \gls{ra} of approximately $10^{\mathrm{h}}27^{\mathrm{m}}50^{\mathrm{s}}.3$.
    Additionally, two polygonal regions around the jet originating from the southern nucleus were excluded based on enhanced velocity dispersions along linear features extending roughly north-south of the southern disc.
    These regions have not been excluded from the cloud-finding analysis.
}.
This comparison also highlights the complementary nature of these two analyses.
The pixel-based method removes the requirement of choosing what conditions indicate boundaries between clouds, and therefore eliminates differences between analyses originating from the chosen definition of a cloud.
However, the pixel-based analysis assumes that each beam is filled with roughly one \gls{gmc} that is the same size as the beam.
The cloud-finding analysis identifies relevant physical size scales so that properties like average surface density, mass, virial parameter, and internal pressure can be calculated using sizes determined from the data.
Also, the ability of the cloud-finding decomposition to automatically analyse lines of sight made up of multiple spectral components allows us to potentially extract more information from the observations.

To facilitate comparison with \citet{Sun2018}, \citet{Bru2021} calculated percentiles and Gaussian \glspl{kde} of the pixel-based distributions using mass weighting, while Figures~\ref{fig:2d_kdes} through \ref{fig:mvir_vs_mco} in this paper use uniform weights for all clouds.
We compared pixel and cloud distributions with both mass and uniform weighting and found that the comparisons were similar in both cases.
Figure~\ref{fig:pixel_cloud_kdes} shows the mass-weighted Gaussian \glspl{kde} comparing several properties of the molecular gas estimated with the pixel and cloud-based methods.

Qualitatively, the distributions of mass surface density and internal pressure are quite consistent between the pixel and cloud-based analyses.
Distributions of peak brightness temperatures from the cloud identification peak at higher values than those of the pixel-based analysis, and the clouds reach somewhat higher temperatures.
Conversely, velocity dispersion, virial parameter, and free-fall time distributions peak at lower values from the cloud measurements than pixels.
Velocity-dispersions measured in clouds are more consistent with pixels that exclude the nuclei (i.e. pixels whose distances from both nuclei are \SI{>1}{\kilo\parsec} after accounting for the inclination angles of the nuclei).
The mass-weighted inner \nth{68} percentiles for distributions from both methods are reported in Table~\ref{tab:pixel_cloud_percentiles}.

\begin{table}
    \centering
    \caption{
        Mass-weighted inner \nth{68} percentiles for the cloud and pixel-based distributions. 
    }
    \begin{threeparttable}
        \label{tab:pixel_cloud_percentiles}
        \begin{tabular}{
            @{}
            c
            S[table-format=3.0]
            S[table-format=4.0]
            S[table-format=3.0]
            S[table-format=4.0]
            S[table-format=3.0]
            S[table-format=4.0]
            @{}
        }
            \toprule
            Quantity                                  & \multicolumn{2}{c}{Cloud} & \multicolumn{2}{c}{Nuclear pixels} & \multicolumn{2}{c}{Non-nuclear pixels} \\
            (units)                                   & {$P_{16}$} & {$P_{84}$}   & {$P_{16}$} & {$P_{84}$}            & {$P_{16}$} & {$P_{84}$} \\
            \midrule
            $\Sigma$                                  & 360        & 2800         & 500        & 2400                  & 140        & 1200 \\
            (\si{\solarmass\per\square\parsec}) \\
            $\sigma_{v}$                              & 21         & 40           & 45         & 100                   & 14         & 47 \\
            (\si{\kilo\metre\per\second}) \\
            $T_{\mathrm{B,max}}$                      & 7          & 20           & 5          & 16                    & 2          & 10 \\
            (\si{\kelvin}) \\
            $\alpha_{\mathrm{vir}}$                   & 2          & 6            & 5          & 34                    & 3          & 15 \\
            \\
            $P_{\mathrm{int}} / k_{B}$                & 7          & 120          & 75         & 720                   & 1          & 93 \\
            (\SI{1e6}{\kelvin\per\cubic\centi\metre}) \\
            $t_{\mathrm{ff}}$                         & 2          & 5            & 3          & 11                    & 3          & 13 \\
            (\si{\mega\year}) \\
            \bottomrule
        \end{tabular}
    \end{threeparttable}
\end{table}

The fact that the velocity-dispersion distributions from non-nuclear pixels and clouds are so similar despite the western region not being included in the pixel distributions likely indicates the spectral decomposition of clouds was successful at separating components along the line of sight.
The multiple spectral components likely arise from independent gas features along the line of sight (e.g. gas from the two progenitor galaxies overlapping in projection) rather than gas with significantly different turbulent motions.
The automatic spectral decomposition provided by cloud finding in the full \gls{ppv} cube is especially useful in morphologically disturbed merger systems where the line of sight can be very complex due to gas overlapping in projection.

While the pixel and cloud distributions are generally consistent, some of the larger differences appear to be related to limitations in the pixel-based analysis.
For example, the higher peak brightness temperatures measured in the cloud distribution are mainly found in regions associated with the jet in the cloud analysis that are not in the pixel-based analysis.
The higher peak temperatures were excluded from the pixel method to ensure a thorough removal of the jets, but that exclusion could have been too aggressive, allowing the cloud analysis to retain more pixels closer to the jet in spatial projection but spectroscopically distinct from it.
Related to the fixed pixel size, the distribution of cloud pressures does not extend as high as estimated in individual pixels.
The internal pressure goes as $P_{\mathrm{int}} \propto R^{-1} \Sigma_{\mathrm{mol}} \sigma_{v}^{2}$ where $R$ is the pixel size in the pixel-based analysis \citep{Sun2018} or the three-dimensional radius for clouds \citep{Ros2021}.
Thus, the constant pixel size appears to overestimate the pressures for the largest clouds more than it underestimates it for the smallest clouds.
Although some care needs to be taken when comparing the extrema of distributions between the pixel and cloud-based methods, if the beam size is close to the median cloud radius then the bulk of the molecular gas properties appear to agree between the two methods.

\subsection{Mass function comparisons} \label{sec:mass_function_discussion}
\subsubsection{Previous cloud decomposition}
The observations presented here have also been independently imaged and analysed through cloud finding with \textsc{cpropstoo} by \citet{Mok2020}.
We adjust the masses from \citet{Mok2020} to use our choices of distance to \ngc{}, \gls{co} \num{2}--\num{1} to \num{1}--\num{0} ratio, and \gls{co}-to-H$_{2}$ conversion factor.
Unlike in our analysis, \citet{Mok2020} do not convolve the observations, retaining the full resolution they achieve of $95 \times \SI{60}{\parsec}$ (adjusted from their adopted distance to \ngc{} of \SI{36}{\mega\parsec} to \SI{44}{\mega\parsec} used here) or about \num{70} per cent the beam area of our \SI{90}{\parsec} \gls{fwhm} beam.
Their noise \gls{rms} is also likely different than ours, because of the different beam sizes and the addition of noise in our data cube to homogenize the \gls{rms} throughout the cube.
Despite these differences, they found \num{123} \glspl{gmc} above their completeness limit of \SI{1.5e7}{\solarmass} where we found \num{120} above that limit (all resolved spatially and spectrally).
Their completeness limit mass is also near where our differential and cumulative mass functions both deviate from high-mass power laws.
Finally, their maximum cloud mass of \SI{2.9e8}{\solarmass} is about \num{20} per cent higher than ours.
% Converting mass from \citet{Mok2020} to our distance, (2-1) to (1-0) ratio, and \gls{co}-to-H$_{2}$ conversion factor:
% M = M_{\mathrm{M}} r_{21,\mathrm{M}} (\alpha_{\mathrm{CO,M}}^{-1} \alpha_{\mathrm{CO}}) (d_{\mathrm{M}}^{-1} d)^{2} \approx 0.38 M_{\mathrm{M}}
% where
%  - $M_{\mathrm{M}}$ is the mass reported by Mok et al. (2020)
%  - $r_{21,\mathrm{M}} = 0.8$ is the \gls{co} 2-1 to 1-0 ratio used by by Mok et al. (2020)
%  - $\alpha_{\mathrm{CO,M}} = \SI{4.35}{\solarmass\per\square\parsec}(\si{\kelvin\kilo\metre\per\second})^{-1}$ is the \gls{co}-to-H$_{2}$ conversion factor used by Mok et al. (2020) and since this is the same as as the 1-0 conversion factor from \citet{Sun2018} who state it accounts for a contribution from helium we assume this does too
%  - $\alpha_{\mathrm{CO}}$ = \SI{1.38}{\solarmass\per\square\parsec}(\si{\kelvin\kilo\metre\per\second})^{-1}$ is our choice of \gls{co}-to-H$_{2}$ conversion factor (including a contribution from helium)
%  - $d_{\mathrm{M}} = \SI{36}{\mega\parsec}$ is the distance to \ngc{} adopted by Mok et al. (2020)
%  - $d = \SI{44}{\mega\parsec}$ is our adopted distance to \ngc{}

To characterize the shape of the \gls{gmc} mass function in \ngc{}, \citet{Mok2020} fit both a Schechter function and pure power law.
The Schechter function results do not place strong constraints on the characteristic cutoff mass and there is not a strong preference for either the Schechter or power law functional form.
\citet{Mok2020} report a pure power law index of \num{-2.10} (with a $1\sigma$ confidence interval from \numrange{-2.20}{-2.00}), which is almost exactly halfway between our low and high mass indices.
Since their low-mass cutoff in fitting the mass function is about a factor of two smaller than our break mass, we would expect their slope to be intermediate between our low and high-mass slopes as it tries to account for some of the curvature in the mass distribution.

Simulated and actual observations of star forming regions analysed at different angular resolutions, with beam-area ranges of about a factor of \num{100}, show very little or no variation in the high-mass slope of source mass functions \citep{Rei2010, Lou2021}.
Therefore, our mass function may be more consistent with that from \citet{Mok2020} than initially anticipated from the differing resolutions, especially considering how close the beam sizes in these analyses of \ngc{} are.
Less easy to predict is the effect of the differing noise levels between the analyses due to the noise level varying throughout the cube used by \citet{Mok2020}.
While \citet{Rei2010} generally found that increasing the noise at fixed resolution resulted in somewhat shallower high-mass slopes in their mass-function fits, the noise difference in the two analyses of \ngc{} depends on where in the cube (position and velocity) each cloud was found.
Since most of our clouds were found near the central velocity of the galaxy, where the original noise is worst, it seems reasonable that the resulting mass function slopes would not be wildly different.

\subsubsection{PHANGS-ALMA}
\citet{Ros2021} performed similar fits to \citet{Mok2020} on the \gls{gmc} mass functions from their sample of \num{10} galaxies, with the addition of attempting to also fit for the effect of completeness causing the turnover at low masses.
Four of the mass functions show significant preference for Schechter-function fits over a pure power law (\ngc{628}, \numlist{2903; 3521; 3627}) but the remainder do not show evidence for a preference.
The pure power-law indices are shown in Figure~\ref{fig:differential_mass_function} as the straight lines overplotted on our differential mass function from \ngc{}, ranging from \numrange{-2.2}{-3.7}.
Our high-mass slope is near the middle of the distribution of slopes from \gls{phangs}.
It is worth noting that robust comparisons over the same mass range cannot be made between \ngc{} and the \gls{phangs} galaxies due to the majority of our most-complete clouds being at or above the most-massive \gls{phangs} clouds.

Some of the same limitations are present in our mass function for \ngc{} as those from \gls{phangs}.
Blending of sources in crowded regions due to coarse spatial resolution will alter the measured mass function shape from the true underlying distribution, and this effect is likely worse in \ngc{} than in the galaxies from \gls{phangs}.
The result of source blending on the shape of a mass function can be difficult to predict since small but not necessarily low-mass clouds will be most blended, blending is worst for the higher-mass clouds for which crowding is worst, and low-mass clouds can be artificially formed by the combination of neighbouring noise peaks or low-mass clouds below the detection threshold \citep{Rei2010}.
Also, \citet{Ros2021} note that their estimates of mass completeness indicate they are only able to measure the mass-function shape over a relatively small mass range (a factor of about ten between the lowest and highest robust masses).
Our rough estimates of where the mass completeness is significantly impacting the shape of the mass function in \ngc{} are where the differential mass function begins to turn over (\SI{\sim 1e7}{\solarmass}) and the double power law break mass (\SI{3e7}{\solarmass}).
Thus, we are also limited to a mass range of about a factor of ten where our mass function is most robust.

Again, these comparisons must be made with caution due to the differing noise levels in our observations of \ngc{} and those from \gls{phangs}.
A lack of clouds in our sample with similar luminosities or masses to \gls{phangs} is predominantly an observational effect, such that the full distribution of clouds in \ngc{} likely includes low-luminosity or mass \gls{phangs}-like clouds.
However, it is not straightforward to predict how improved-sensitivity observations of \ngc{} would alter the cloud distribution.
First, we would detect gas down to lower surface densities, which would reveal new low-mass clouds as well as add low surface-density gas to the outer extents of already-identified clouds.
Whether this new gas is assigned to new or pre-existing clouds would depend on the \gls{s/n} contrast of the new emission.
The second effect is the tendency for algorithms like \textsc{pycprops} to break emission up into roughly beam-sized structures, so the spatial extent of the new emission would also impact how it was assigned to new or existing clouds.
It is clear, though, that a population of clouds on scales of \SI{90}{\parsec} in \ngc{} is significantly more luminous (and likely also more massive) than in the \gls{phangs} galaxies.

\section{Conclusions} \label{sec:conclusions}
We have performed molecular-cloud identification on observations of the nearest \gls{lirg}, \ngc{}, at a matched resolution of \SI{90}{\parsec} to the \gls{phangs} cloud-finding results presented by \citet{Ros2021}.
In these \gls{co} (\num{2}--\num{1}) observations we have identified \num{185} spatially as well as spectrally resolved clouds, which in almost all properties analysed are extreme relative to the \gls{phangs} sample.
Our full catalogue of clouds is provided in Tables~\ref{tab:cloud_properties_1} and \ref{tab:cloud_properties_2}.

Cloud velocity dispersions, luminosities, \gls{co}-estimated masses, mass surface densities, virial masses, virial parameters, size-linewidth coefficients, and internal turbulent pressures are all significantly higher than values measured in clouds in the \gls{phangs} galaxies.
Radii are slightly larger in \ngc{} and free-fall times slightly shorter.
However, the distribution of cloud eccentricities measured in \ngc{} is often indistinguishable from those from the \gls{phangs} sample.
Explanations for the similarities and differences across these properties are discussed in Section~\ref{sec:discussion}.

% - high internal pressures are probably mostly balanced by high external pressures
%   - high velocity dispersions but not very different radii
%   - there must be bound and collapsing gas since the \gls{sfr} is so high in 3256
%   - may still be excess turbulence making the radii a bit larger
%   - also the broadest distribution of internal pressures
% - narrower distribution of virial parameters
%   - maybe narrower range of dynamical states in ngc 3256
%   - but eccentricities may indicate that the average cloud dynamical state is similar between ngc and phangs i.e. clouds in 3256 may be mostly bound despite high velocity dispersions and high naive virial parameters
% - free-fall times are similar at our \SI{90}{\parsec} resolution as to the \SI{512}{\parsec} resolution observations analysed by \citet{Wil2019}
%   - implies similar volume densities of molecular gas
%   - would require a smooth molecular \gls{ism} to not see beam dilution changing the free-fall times across such different scales
%   - consistent with signatures of smooth \gls{ism} from \citet{Bru2021}

Despite differences in how the data were prepared, the mass function of clouds in \ngc{} measured here appears roughly consistent in power-law slope with the independent analysis of these observations by \citet{Mok2020}.
Compared to the mass function shapes derived from the \gls{phangs} galaxies by \citet{Ros2021}, the high-mass portion from \ngc{} appears near the middle of their distribution of slopes.

% - High cloud masses consistent with prediction of relatively high-mass star clusters \citep[observed by][]{Ada2020} being formed from more massive clouds.
% - High cloud mass surface densities consistent with prediction of relatively high \gls{cfe} and observed by \citet{Ada2020}.

Comparison of this analysis with a pixel-based approach used by \citet{Bru2021} shows general agreement between the measured molecular-gas properties.
Cloud and pixel-based analyses appear to be complementary in this case as the pixel analysis naturally does not require choosing what defines the edge of a ``cloud'', while the cloud analysis can include more observed regions because of its ability to decompose multiple spectral components along the line of sight.
Given the median cloud radius found here is \SI{100}{\parsec} and the pixel analysis was performed with resolutions of \SIlist{55; 80; 120}{\parsec}, the pixel analysis was potentially resolving the clouds more than initially expected.
The largest differences between the two methods appear to be related to the limitations imposed by the pixel-based analysis assuming clouds of a fixed size or having to avoid spectrally complex lines of sight.

% Future work
% - longer term for observational improvement of SFR may be JWST observations
% - both the radio continuum and JWST could also give dust masses for independent measurement of molecular-gas mass surface density
%    - independent of alpha_CO
%    - but would need to know dust-to-gas ratio (GDR)
%    - or maybe you could solve for alpha_CO, GDR, and mass like Sandstrom et al. (2013)?

Future observations of \gls{co} at even higher-resolution could begin to probe the gas properties within individual clouds.
We could first search for signatures of the \gls{ism} becoming clumpier at resolutions below \SI{55}{\parsec}.
Given the high \gls{sfr} in \ngc{} there must eventually be a scale where the molecular gas decouples from its surroundings and would be observed to be collapsing.
Observing gas within individual clouds would also likely reveal the small-scale properties that set the self-gravitating threshold density of the gas as well as the fraction of self-gravitating gas.

\section*{Acknowledgements}
We are grateful to Dr. E. Rosolowsky for many helpful discussions about the details of how \textsc{pycprops} works and on ways to match our analysis to that of \gls{phangs}.
We are also grateful to Dr. G. Eadie for helpful discussion about logistic regression.
We thank the anonymous referee for detailed comments that improved the content of this paper.

The research of NB is partially supported by the New Technologies for Canadian Observatories, an NSERC CREATE program.
The research of CDW is supported by grants from the Natural Sciences and Engineering Research Council of Canada and the Canada Research Chairs program.

We acknowledge the use of the ARCADE (ALMA Reduction in the CANFAR Data Environment) science platform.
ARCADE is a ALMA Cycle 7 development study with support from the National Radio Astronomy Observatory, the North American ALMA Science Centre, and the National Research Centre of Canada.

This research made use of \textsc{Astropy}, a community-developed core \textsc{Python} package for Astronomy \citep[\url{http://www.astropy.org},][]{astropy2013,astropy2018}.
This research also made use of the \textsc{SciPy} \citep{Vir2020}, \textsc{Matplotlib} \citep{Hun2007}, \textsc{NumPy} \citep{van2011}, \textsc{pandas} \citep{McK2010}, \textsc{scikit-learn} \citep{Ped2011}, \textsc{Jupyter Notebook} \citep{Klu2016}, and \textsc{statsmodels} \citep{Sea2010} \textsc{Python} packages.
This research has made use of the \gls{carta} \citep{Com2021}.
This research has made use of NASA’s Astrophysics Data System.
This research has made use of the VizieR catalogue access tool \citep{Och2000}.
This research has made use of the NASA/IPAC Extragalactic Database (NED), which is funded by the National Aeronautics and Space Administration and operated by the California Institute of Technology.
This research has made use of the SIMBAD database, operated at CDS, Strasbourg, France \citep{Wen2000}.

\section*{Data Availability}
This paper makes use of the following ALMA data: ADS/JAO.ALMA\#2015.1.00714.S (accessed from the \gls{alma} Science portal at \url{almascience.org}).
ALMA is a partnership of ESO (representing its member states), NSF (USA) and NINS (Japan), together with NRC (Canada), MOST and ASIAA (Taiwan), and KASI (Republic of Korea), in cooperation with the Republic of Chile.
The Joint ALMA Observatory is operated by ESO, AUI/NRAO and NAOJ.
The National Radio Astronomy Observatory is a facility of the National Science Foundation operated under cooperative agreement by Associated Universities, Inc.

The derived data generated in this research will be shared on reasonable request to the corresponding author.

%%%%%%%%%%%%%%%%%%%%%%%%%%%%%%%%%%%%%%%%%%%%%%%%



\bibliographystyle{mnras}
\bibliography{references}

%%%%%%%%%%%%%%%%%%%%%%%%%%%%%%%%%%%%%%%%%%%%%%%

%%%%%%%%%%%%%%%%% APPENDICES %%%%%%%%%%%%%%%%%%%%

\appendix

\section{Effect of pixel size on velocity-dispersion distributions} \label{sec:dispersion_pixel_size}
Since the spatial pixel size of interferometric cubes is a free parameter during imaging and cleaning, we originally chose to have about five pixels across the native beam \gls{fwhm} of \SI{\approx 47}{\parsec} to safely oversample each resolution element.
Cubes were then convolved to have a range of larger beam sizes, and before the pixel-based analysis by \citet{Bru2021} the cubes were regridded to have two pixels across each beam \gls{fwhm} to roughly Nyquist sample each resolution element.
When trying to decide what pixel size to use for this cloud-finding analysis, we compared results with a range of pixel sizes from \num{2} to almost \num{14} pixels across the beam \gls{fwhm} at different spatial resolutions.
We initially intended to use the cube with two pixels across the beam to make comparisons with our previous pixel analysis more direct, but we found the distribution of velocity dispersions changed with the choice of pixel size, regardless of spatial resolution.

\begin{figure}
    \centering
    \includegraphics{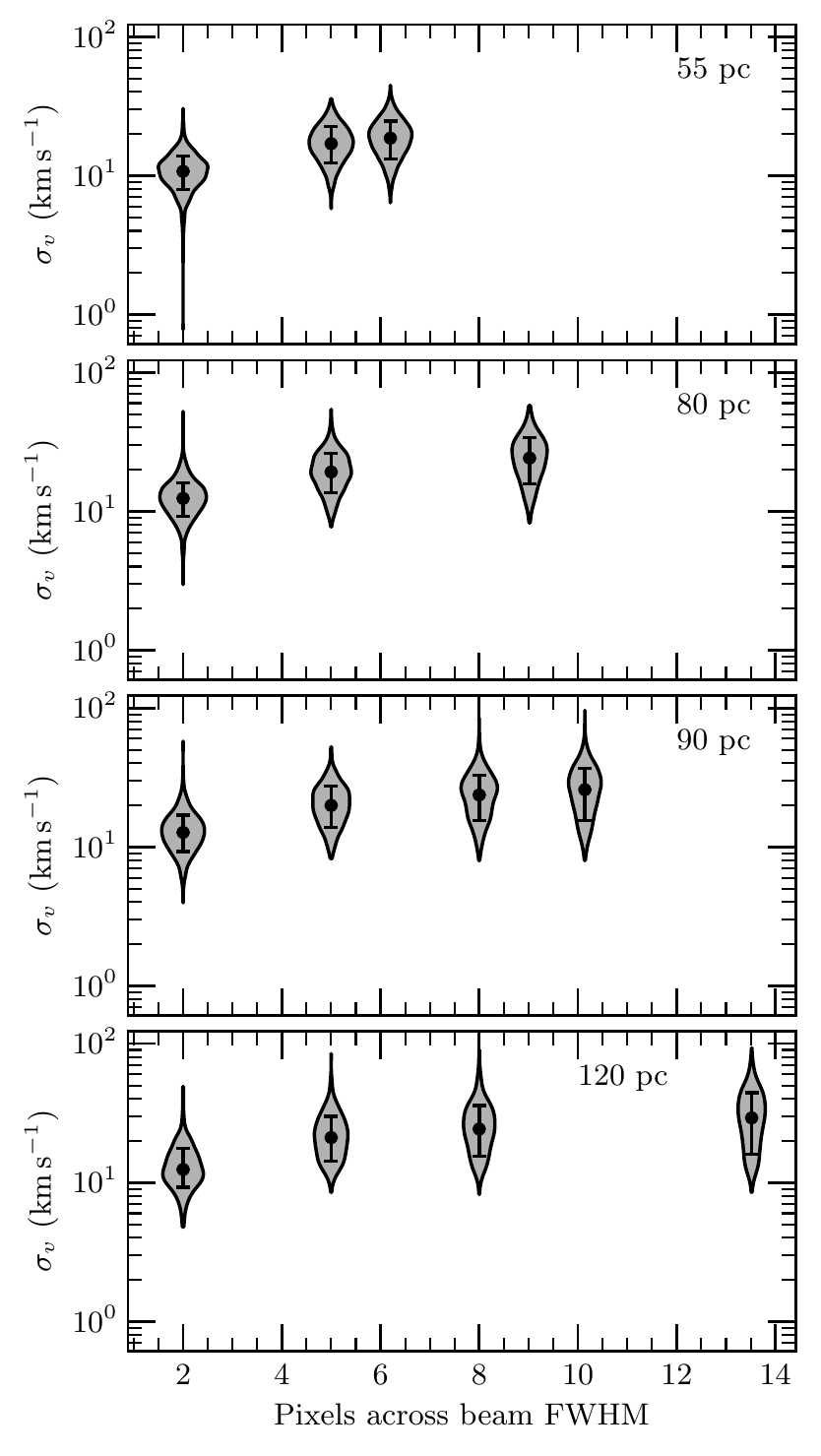}
    \caption{
        Distributions of cloud velocity dispersions vs. pixel size (measured in the number of pixels across the beam \gls{fwhm}) with results at different spatial resolutions shown in each row.
        Circles show medians and errorbars show \nth{16} to \nth{84} percentile ranges.
        Curves and shaded regions show Gaussian \glspl{kde}, with uniform weights for all clouds and widths normalized across all distributions to show relative cloud fractions per dex.
    }
    \label{fig:dispersion_vs_pixel_size}
\end{figure}

Figure~\ref{fig:dispersion_vs_pixel_size} shows how the cloud velocity-dispersion distributions, medians, and inner \nth{68} percentile ranges change with pixel size at a range of spatial resolutions.
Two and five pixels across each beam \gls{fwhm} were tested, along with eight if possible, and finally the smallest pixels came from not regridding the cubes after imaging with the native \SI{47}{\parsec} beam.
Across all resolutions, the medians are lower for larger pixels (e.g. fewer pixels across the beam \gls{fwhm}) and potentially show signs of convergence towards the smallest pixels.
These cloud-finding tests were carried out on cubes that did not have homogenized noise so it is best to compare the distributions at a single spatial resolution to avoid lower noise levels at poorer resolution influencing the cloud properties.
Between pixel sizes, the standard error on the mean noise \gls{rms} was at most half a per cent of the median \gls{rms}, so the changes in the distributions are not likely driven by the changing noise level.

\begin{figure*}
    \centering
    \includegraphics{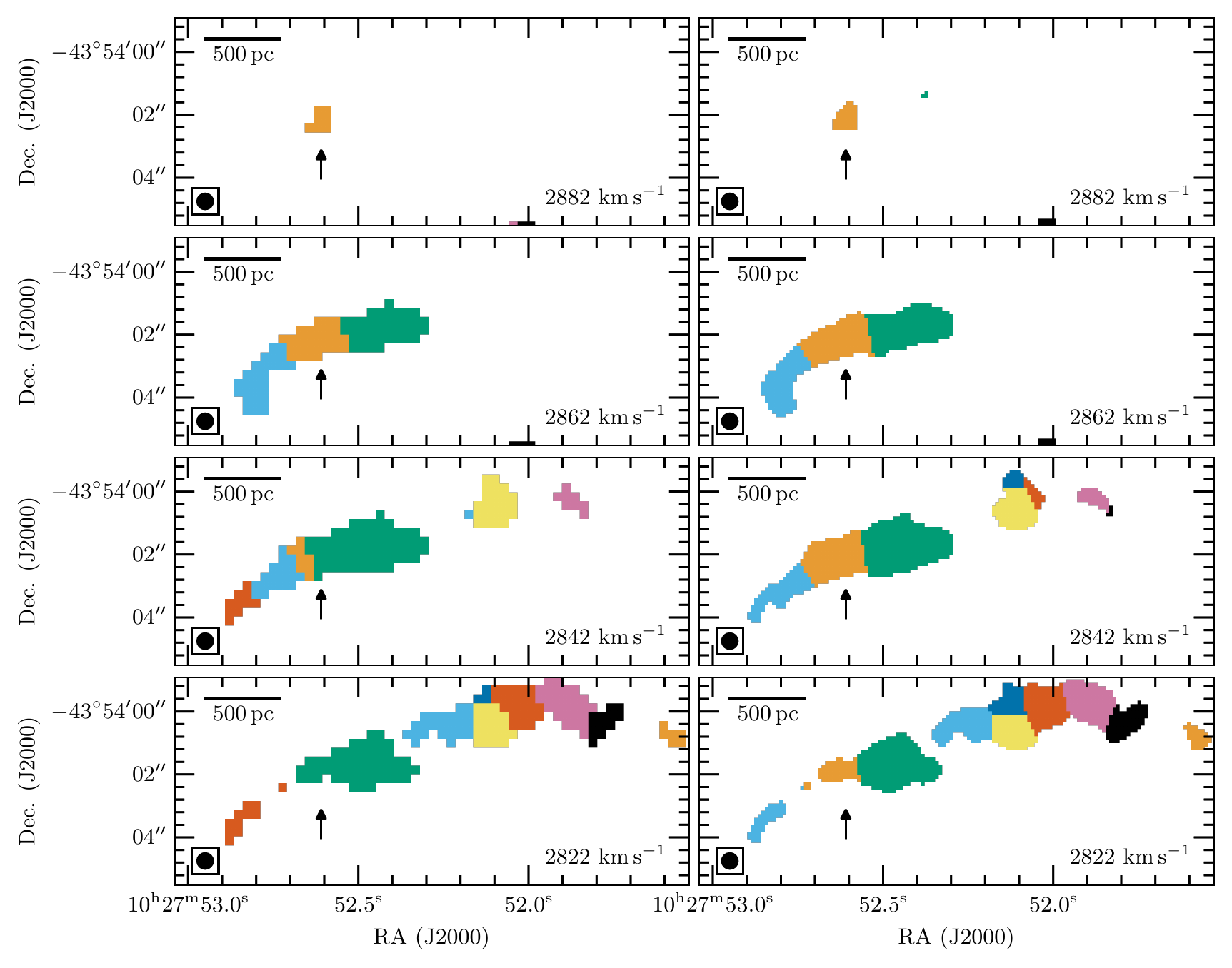}
    \caption{
        Spatial maps at \SI{120}{\parsec} resolution showing identified clouds labelled by colour and at different velocity slices (rows) when the emission cube was regridded to have two (left) and five (right) pixels across the beam \gls{fwhm}.
        Velocities of the slices are shown in the bottom-right of each panel, the beam size is shown as the black circles in the bottom-left, and a scale bar in the top-left indicates \SI{500}{\parsec} at the distance of \protect\ngc{}.
        Arrows in each panel highlight a cloud that with large pixels has the adjacent cloud to the right ``wrap around'' it along the velocity axis so its velocity range is reduced, but with small pixels both clouds persist along the same range in velocity.
        Also note how the cloud to the left of the arrow splits into two clouds across this velocity range with large pixels but is identified as a single cloud throughout with small pixels.
    }
    \label{fig:cloud_labels}
\end{figure*}

Visual inspection of the cloud boundaries suggests that the change in velocity-dispersion distributions was because the boundaries between clouds changed with changing pixel size.
Figure~\ref{fig:cloud_labels} shows a small grouping of clouds to the north of the nuclei along a spiral-arm feature, where the emission was relatively isolated so the cloud boundaries should be easiest to follow through the cube.
The smallest pixels result in some clouds remaining spatially distinct over more velocity channels.
In contrast, larger pixels sometimes resulted in the boundary between two clouds shifting in the spatial dimensions while moving along the spectral dimension.
These shifting boundaries resulted in the emission at one spatial location being split between two clouds along the velocity dimension, so that one cloud ``wrapped around'' a spatially adjacent one in velocity.
This wrapping scenario is visible for the cloud marked with an arrow in Figure~\ref{fig:cloud_labels}, where the arrow is drawn at the same location at each velocity slice and at both pixel sizes.

Also visible in Figure~\ref{fig:cloud_labels} is how the cloud to the left of the arrow splits into two clouds by the \SI{2842}{\kilo\metre\per\second} panel with large pixels but remains a single cloud at all velocities with smaller pixels.
A noise peak/trough combined with large pixels could result in a larger gradient, causing the emission to be split into more than one cloud.
However, smaller pixels may better resolve the same feature as a smooth gradient along the cloud so it does not split.
Both the wrapping and splitting along the velocity axis in the 2 pixel version will result in lower velocity dispersions since the clouds span fewer channels.

Given the main point of comparison in this work is to the \gls{phangs} cloud-finding results from \citet{Ros2021}, we chose to match spatial resolution by carrying out the remainder of our analysis on the \SI{90}{\parsec} resolution cube.
As for pixel size, we chose to use the smallest pixels we could with a cube that was derived from the same imaging by \citet{Bru2021}.
Changes in the velocity-dispersion distributions, medians, and inner \nth{68} percentile ranges decrease as the number of pixels across the beams increase, possibly indicating the measurements with small pixels are closer to converged/``true'' values.
Smaller pixels also have the benefit of reducing the amount of velocity wrapping and splitting of the found clouds.

\section{Additional scaling-relation views} \label{sec:more_scaling_relations}
\begin{figure*}
\centering
\includegraphics{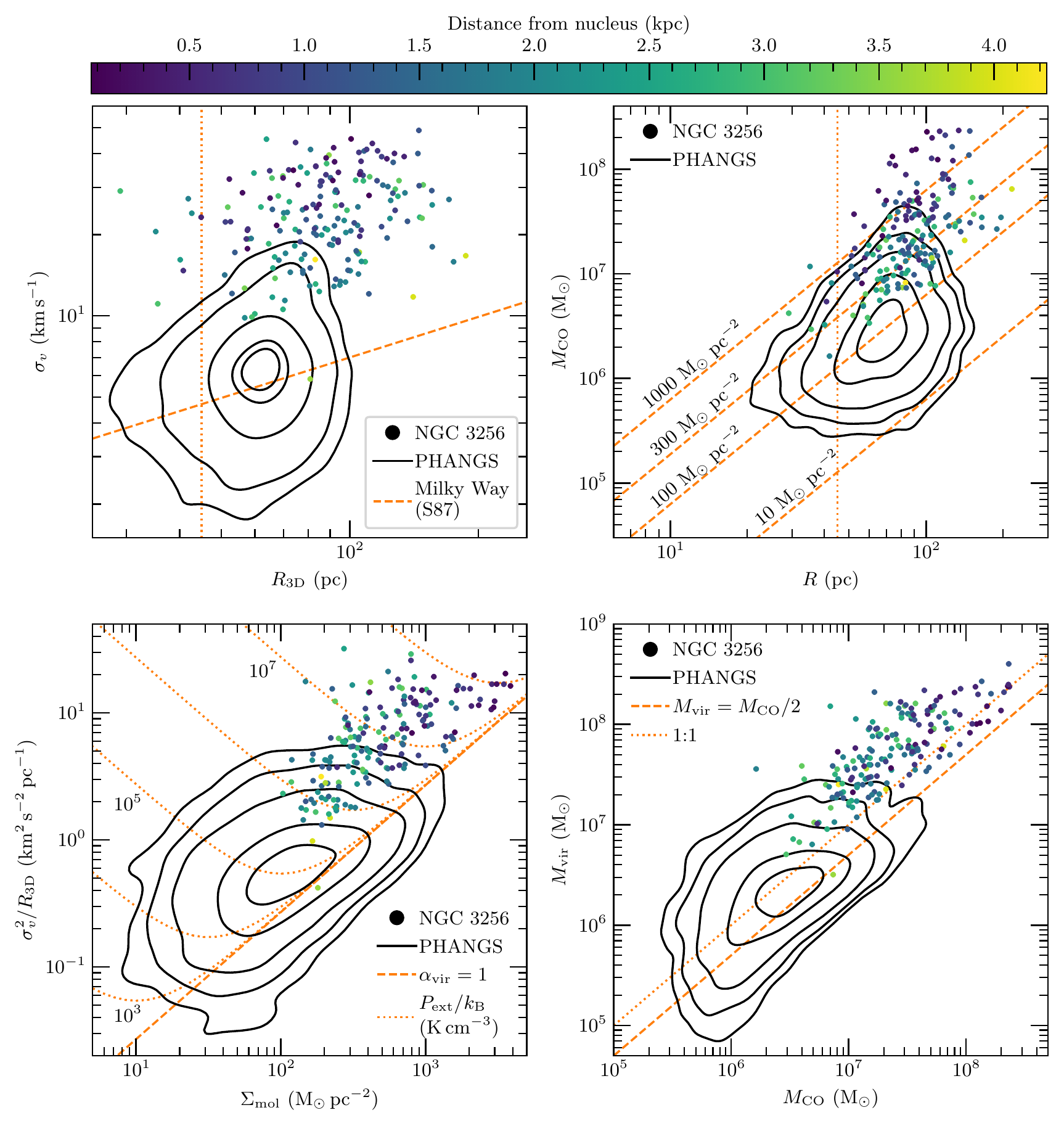}
\caption{
    Same as Figures~\ref{fig:vd_vs_3d_radius} through \ref{fig:mvir_vs_mco} but now the points from the \protect\ngc{} clouds are coloured by their distance from one of the progenitor nuclei.
    The smaller of the two distances from the northern or southern nucleus is shown.
    Positions of the nuclei are \gls{ra} $10^{\mathrm{h}}27^{\mathrm{m}}51^{\mathrm{s}}.226$ \gls{dec} \ang{-43;54;13.942} for the northern nucleus and \gls{ra} $10^{\mathrm{h}}27^{\mathrm{m}}51^{\mathrm{s}}.221$ \gls{dec} \ang{-43;54;19.168} for the southern nucleus.
}
\label{fig:distance_color_scalings}
\end{figure*}

Presented here are the same scaling relations shown in Figures~\ref{fig:vd_vs_3d_radius} through \ref{fig:mvir_vs_mco} but highlighting some additional features.
In Figure~\ref{fig:distance_color_scalings}, the colour of the points from \ngc{} now indicates the distance from the cloud centre to one of the progenitor nuclei.
The distance to both nuclei was calculated for each cloud and the smaller distance is used to colour the points.
Positions of the nuclei are the same as used by \citep{Bru2021} and are \gls{ra} $10^{\mathrm{h}}27^{\mathrm{m}}51^{\mathrm{s}}.226$ \gls{dec} \ang{-43;54;13.942} for the northern nucleus and \gls{ra} $10^{\mathrm{h}}27^{\mathrm{m}}51^{\mathrm{s}}.221$ \gls{dec} \ang{-43;54;19.168} for the southern nucleus.
Generally, clouds with the highest values for any of the properties shown are found closer to the nuclei while most of the rest of the distributions are at a mix of distances.

\begin{figure*}
\centering
\includegraphics{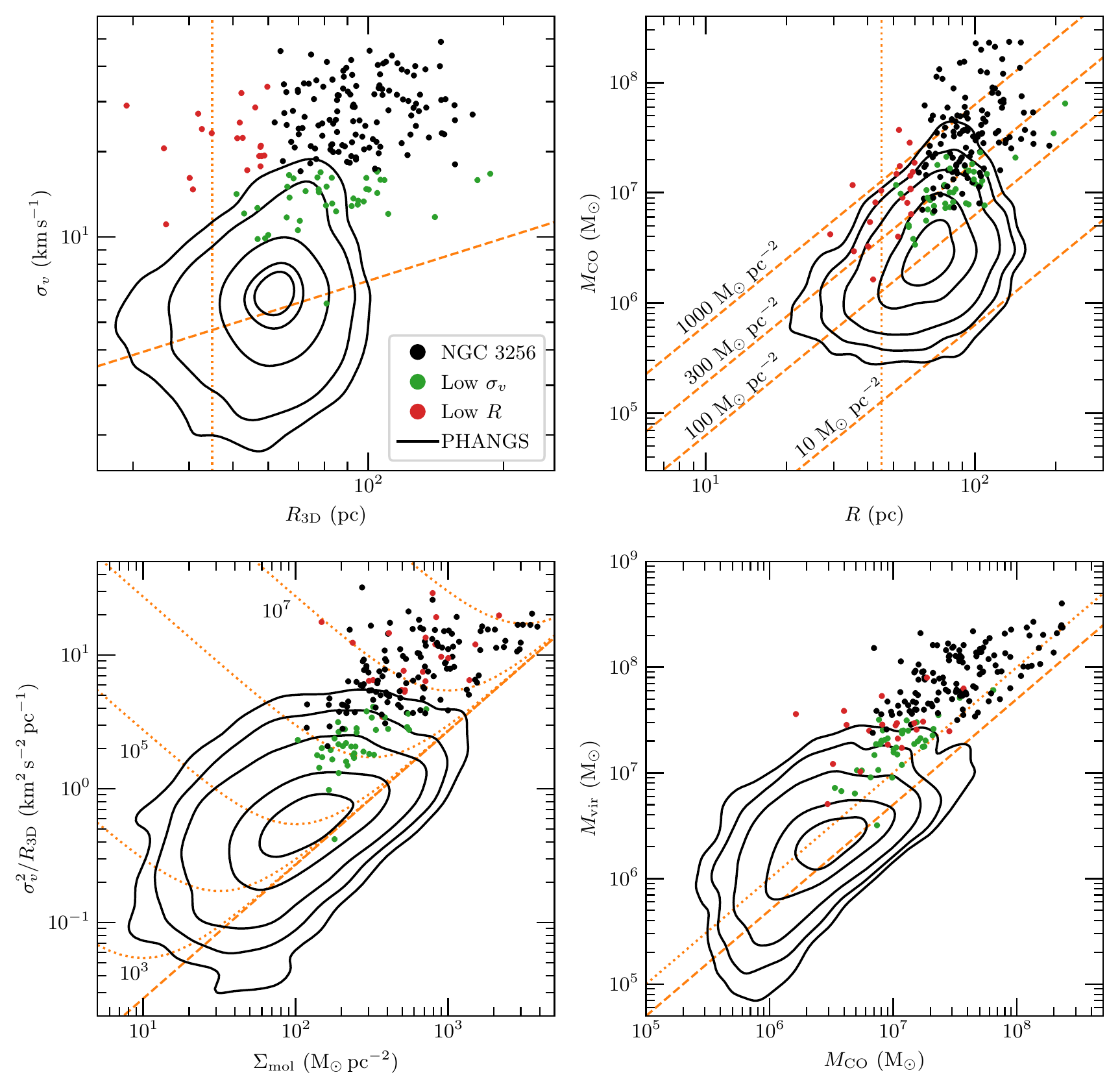}
\caption{
    Same as Figures~\ref{fig:vd_vs_3d_radius} through \ref{fig:mvir_vs_mco} but clouds with small velocity dispersions or radii marked with colour in each panel.
    Green points show clouds with $\sigma_{v} < \SI{17}{\kilo\metre\per\second}$ and red points show clouds with $R < \SI{60}{\parsec}$.
    If a cloud met both of the criteria then the colour was chosen based on which property has the largest per cent difference from the threshold.
}
\label{fig:low_value_scalings}
\end{figure*}

To explore if it is always the same clouds in \ngc{} that are most consistent with \gls{phangs} throughout Figures~\ref{fig:vd_vs_3d_radius} through \ref{fig:mvir_vs_mco}, we have highlighted clouds either with the smallest values of velocity dispersion or two-dimensional radius in Figure~\ref{fig:low_value_scalings}.
The criteria are easiest to see in the top-left panel as clouds that had $\sigma_{v} < \SI{17}{\kilo\metre\per\second}$ are in green and $R < \SI{60}{\parsec}$ are in red.
The colour for clouds that met both of the criteria was chosen by the quantity with the larger per cent difference from the corresponding threshold.
For example, a cloud with a velocity dispersion of \SI{2}{\kilo\metre\per\second} and two-dimensional radius of \SI{40}{\parsec} would be shown in green.
Broadly, those two groups are typically the clouds that are most similar to \gls{phangs} or their trends.
This is not true, however, for small-radius clouds in the velocity dispersion vs. radius and size-linewidth coefficient vs. mass surface density plots, where clouds from \ngc{} appear where there are no clouds from \gls{phangs}.

%If you want to present additional material which would interrupt the flow of the main paper, it can be placed in an Appendix which appears after the list of references.

%%%%%%%%%%%%%%%%%%%%%%%%%%%%%%%%%%%%%%%%%%%%%%%

% Don't change these lines
\bsp	% typesetting comment
\label{lastpage}
\end{document}